\documentclass[review]{elsarticle}

\usepackage[utf8]{inputenc}

\usepackage{lineno}
\modulolinenumbers[5]

\journal{Journal of \LaTeX\ Templates}
\usepackage[colorinlistoftodos]{todonotes}
\usepackage{graphicx}
\usepackage{float}
\usepackage{soul}
\usepackage{array}
\usepackage{tabularx}
\usepackage{changes}
\graphicspath{ {./figures/} }
\usepackage{url}

\usepackage{breakurl}
\usepackage[breaklinks]{hyperref}

\usepackage{subcaption}
\usepackage[labelfont=bf]{caption}

\usepackage[margin=3cm]{geometry}
\usepackage{adjustbox}
\usepackage{multirow}

\begin{document}

\begin{frontmatter}
\title{Assessing the Potential for Building Sector Retrofits to Mitigate ERCOT Electricity Shortfalls During Winter Storm Uri}

\author[add1]{Matthew J. Skiles\corref{cor1}}
\author[add1]{Joshua D. Rhodes, PhD}
\author[add1]{Michael E. Webber, PhD}

\address[add1]{Walker Department of Mechanical Engineering, The University of Texas at Austin, 204 E Dean Keeton St, Austin, TX 78712, USA}
\cortext[cor1]{Corresponding Author.\\ 
\emph{Email Addresses:} mskiles@utexas.edu (Matthew J. Skiles), joshdr@austin.utexas.edu (Joshua D. Rhodes), webber@mail.utexas.edu (Michael E. Webber)}

\begin{abstract}
This analysis investigates energy performance of the residential and commercial building sectors in the Electric Reliability Council of Texas (ERCOT) during Winter Storm Uri. ERCOT electricity demand was modeled for the ERCOT baseline building stock as well as for the baseline building stock retrofitted with an efficiency upgrade package, an electrification upgrade package, and an efficiency + electrification upgrade package. The electrification scenario that retrofitted buildings with air-source heat pumps (ASHPs) would have lowered ERCOT daily peak electricity demand relative to the baseline scenario for every day of the year, except during the week of Winter Storm Uri. As the mean outdoor temperature dropped below -5°C (23°F), diminishing ASHP efficiency would have resulted in electrification scenario demand exceeding the two distinct baseline scenario daily demand peaks on February 15\textsuperscript{th} and 16\textsuperscript{th} (87.3 GW and 88.7 GW) to hit 111.8 GW and 117.5 GW. The efficiency package would have lowered daily peak demand on these days to 67.0 GW and 68.0 GW. The efficiency + electrification package would have lowered peak demand on these days to 81.5 GW and 85.6 GW. When electricity shortfall profiles were produced by comparing modeled electricity demand to actual ERCOT electricity generation during the storm, the results indicate that the electrification scenario electricity shortfall (1741 GWh) would have been larger than for the baseline scenario (1225 GWh) and the electricity shortfalls for the efficiency scenario (347 GWh) and efficiency + electrification scenario (704 GWh) would have been lower than the baseline. The efficiency, electrification, and efficiency + electrification scenarios would all have lowered summer daily peak demand due to improvements in building cooling efficiency and would have lowered annual electricity consumption by 5.9\%, 6.8\%, and 11.9\%, respectively.
\end{abstract}

\begin{keyword} Building Energy Efficiency, Building Electrification, ERCOT, Peak Demand, Energy Security
\end{keyword}
\end{frontmatter}

\pagebreak
\section{Introduction}

The major North American winter storm of February 2021, commonly known as “Winter Storm Uri”, impacted much of the United States. The most pronounced impacts to infrastructure occurred in Texas, where 69\% of Texans lost power for an average of 42 hours \cite{UofH2021}. Disruptions in power supply contributed to a cascading series of infrastructure failures, where blackouts impacting natural gas infrastructure contributed to a nearly 50\% reduction in natural gas production in Texas \cite{EIA2021b}. Subsequent interruptions in natural gas deliveries to power plants forced even more power generation capacity out of service. Power outages at water treatment facilities also led to disruptions in water service for 49\% of Texans \cite{UofH2021}. Unfortunately, this electricity shortfall occurred at the worst possible time, when the Electric Reliability Council of Texas (ERCOT) estimates that system demand peaked at 76,819 MW \cite{king2021}, which if it had been served, would have set a new ERCOT load record. Not only would this have been the highest load ever recorded in ERCOT at the time, it also would have been the first time that winter peak load had surpassed summer peak load in a given year \cite{skiles2023}. By some estimates, Winter Storm Uri resulted in 246 deaths \cite{SHS2021} and 80 billion USD to 130 billion USD in economic losses in Texas \cite{golding2021}. 

In the wake of this disaster, significant attention has been given to implementing policy that could improve ERCOT grid reliability in the face of a similar meteorological event. Most of the solutions have been supply-side interventions intended to improve the reliability of electricity generation and delivery. The Texas state government has adopted new winterization regulations \cite{RCOT2022}, \cite{PUC} designed to mitigate the weather-driven outages that forced up to 48\% of ERCOT generating capacity offline during the storm \cite{clack2021}. Additionally, a new market mechanism, known as the Performance Credit Mechanism, was passed into law which pays generators if they are available to generate electricity when grid conditions are tight \cite{foxhall2023}, \cite{walton2023}. The Texas government has also created the 7.2 billion USD Texas Energy Fund, which provides completion bonuses and low interest loans to developers for building new dispatchable power plants \cite{baskar2023}.

An alternative and largely ignored approach to reducing the ERCOT electricity shortfall during Winter Storm Uri is to lower peak electric load through demand-side interventions. Weather-sensitive residential and commercial sector loads could be a good target for intervention, as they increased dramatically in response to Winter Storm Uri temperatures to comprise up to 83\% of daily peak served load (Figure 1a-b). When ERCOT power generators can generate enough power to meet electricity demand, electricity demand is equivalent to served electricity load. However, during February 2021, ERCOT was forced to request load shed and served electricity load dropped below electricity demand.

\begin{figure*}
\hfill
\begin{minipage}[b]{\textwidth}
\begin{subfigure}{\textwidth}
\subcaption{}
\includegraphics[scale=0.6]{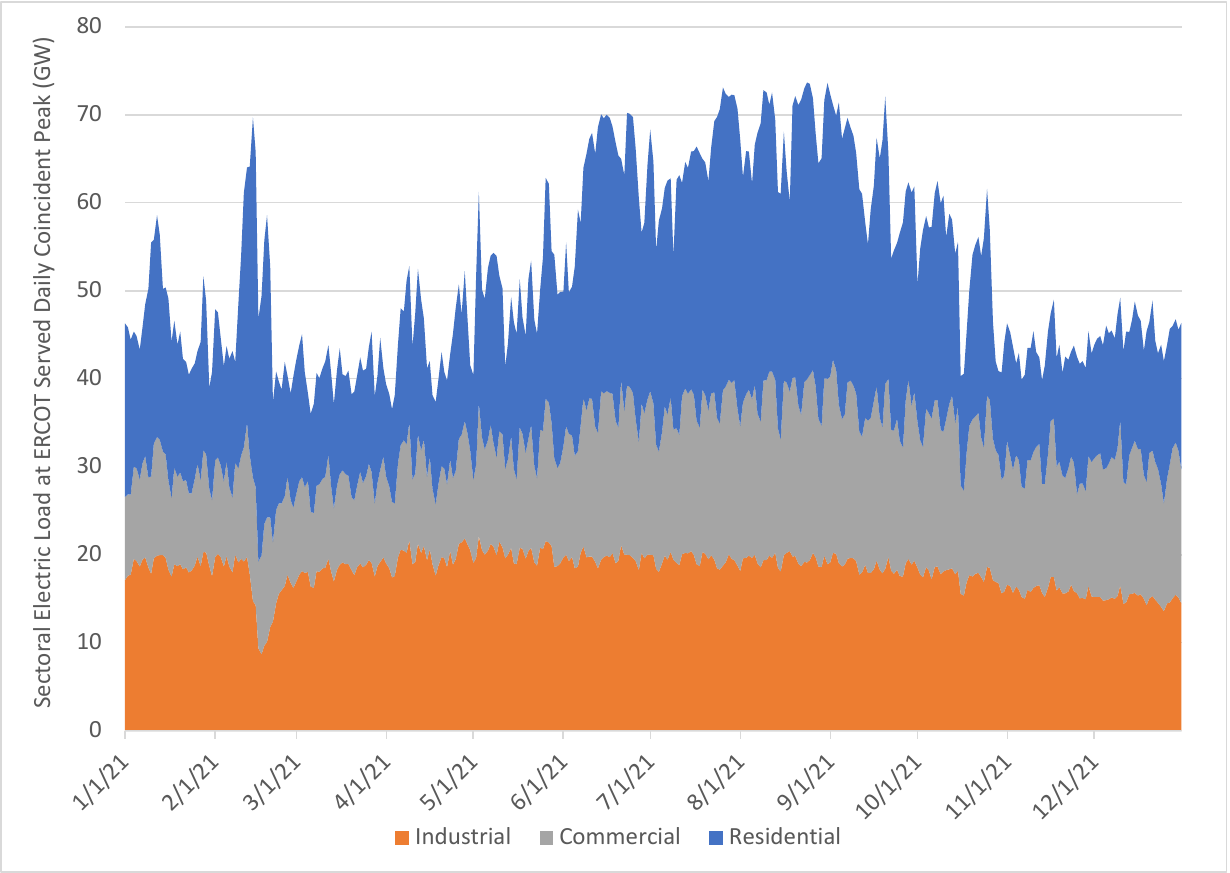}
\centering
\end{subfigure}\par
\begin{subfigure}{\textwidth}
\subcaption{}
\includegraphics[scale=0.6]{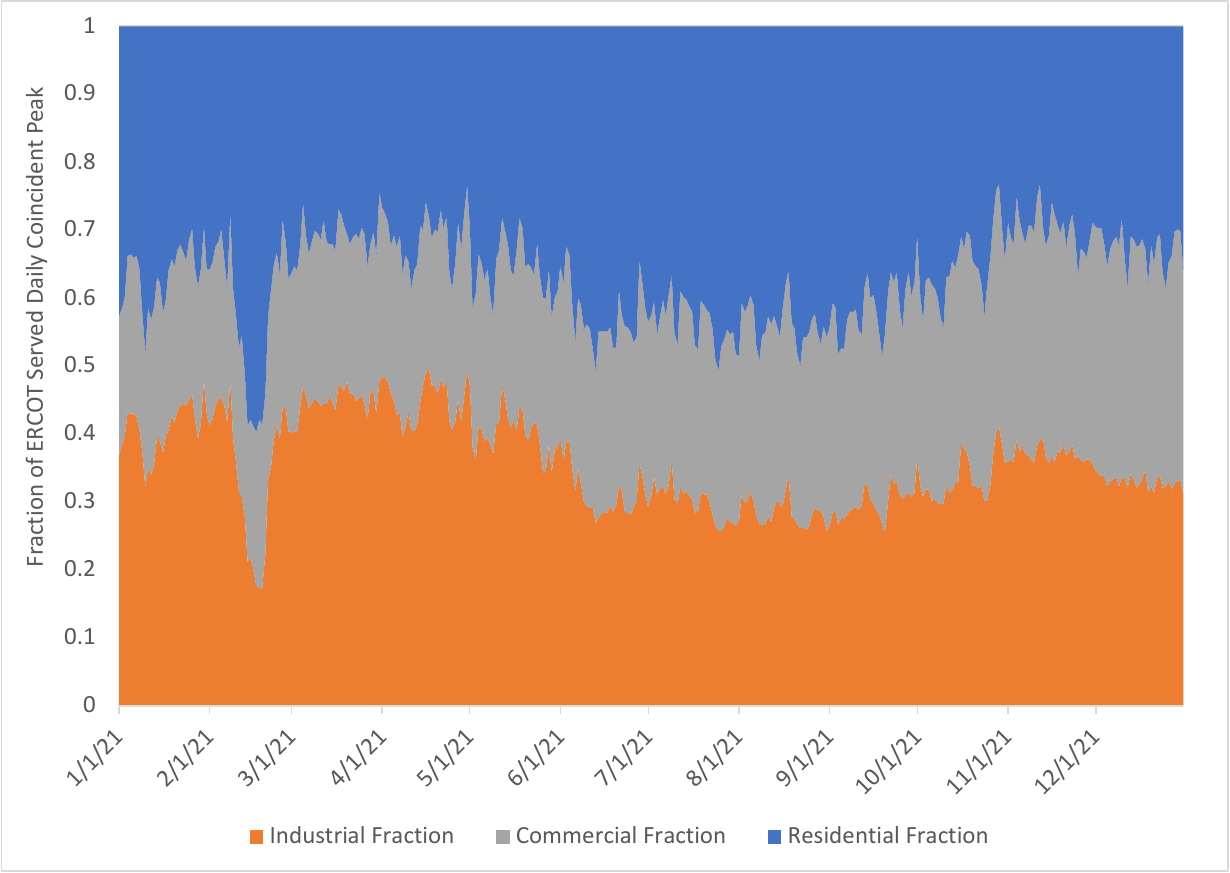}
\centering
\end{subfigure}
\end{minipage}
\quad
\begin{minipage}[b]{\textwidth}
\caption{\textbf{a,} An isolated spike in served residential and commercial electric load during Winter Storm Uri resulted in system-wide daily coincident peak served load that was comparable to daily coincident peak served load during the summer. \textbf{b,} Served residential and commercial sector electric load comprised 50\% – 74\% of daily coincident peak served load outside of February, but increased to 83\% of daily coincident peak served load during Winter Storm Uri. \cite{ERCOT2021c}.}\label{key}
\centering
\end{minipage}
\end{figure*}

\pagebreak

One limitation of served electric load data is that the highest resolution data typically available for the residential and commercial sector is at the building level, and data about specific end-use consumption within buildings is generally not available. Considering the outsized role that building electricity demand played in this event, it is important to characterize building electric load in more detail and explore methods for reducing peak building demand during severe weather events. 

Texas building heating demand is particularly sensitive to cold outdoor temperatures because Texas building envelopes are relatively inefficient. Texas had no mandatory building energy codes until 2001 \cite{EERE}, and survey data indicate that only 14\% of Texas residential buildings have combined roof and ceiling insulation values of at least R-38 \cite{wilson2021}, the minimum residential building ceiling insulation level recommended for Texas climate zones by the International Residential Code \cite{IRC2021}. Only eight other states report lower prevalence of R-38 or greater ceiling and roof insulation in their residential building stock \cite{wilson2021} (Figure 2). 

\begin{figure}[H]
\includegraphics[scale=0.35]{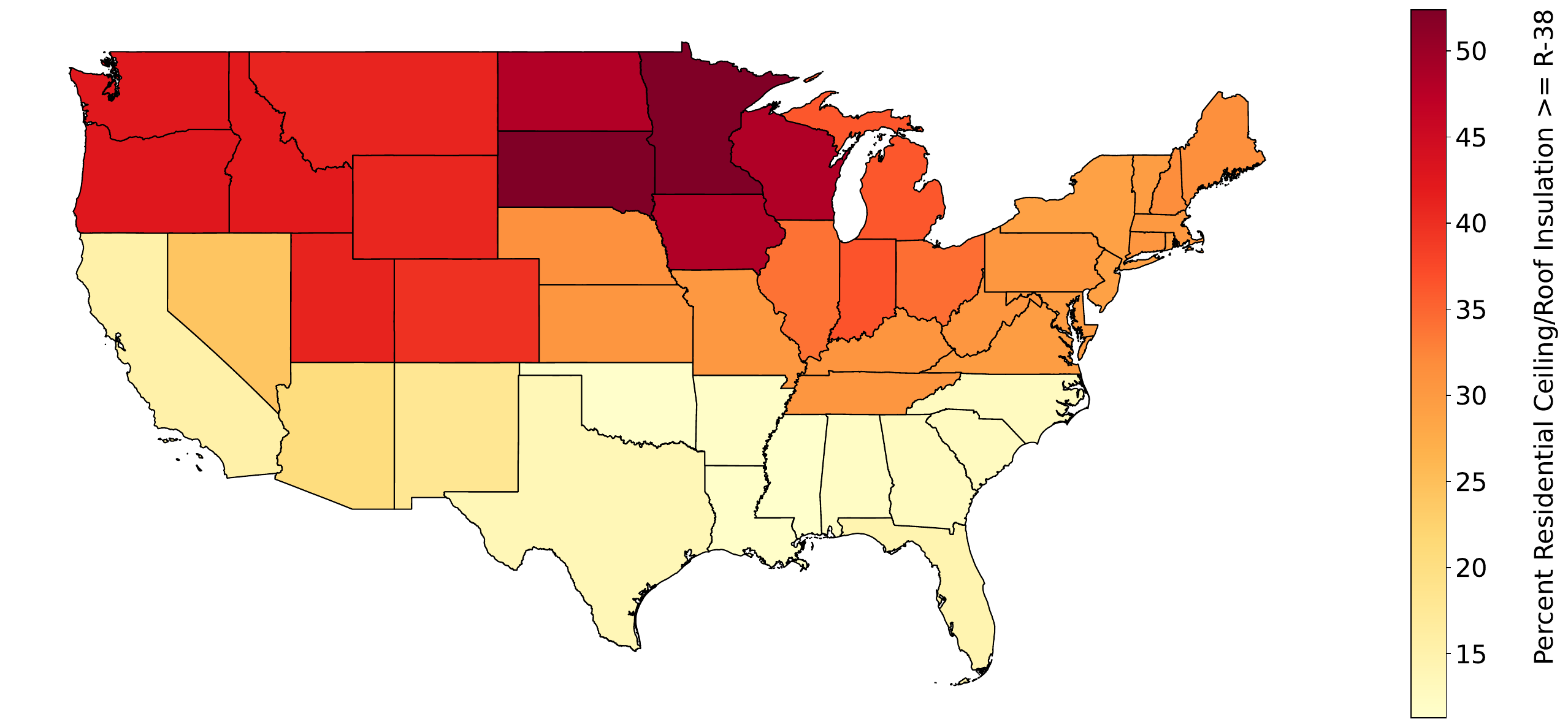}
\caption{Compared to other U.S. states, a relatively low percentage of Texas residential buildings have combined ceiling and roof insulation that meets or exceeds R-38 \cite{wilson2021}, the minimum level of ceiling insulation recommended for Texas climate zones by the International Residential Code \cite{IRC2021}.}
\centering
\end{figure}

The R-value is a measure of the resistance to heat transfer provided by insulation. Higher R-values are better and indicate greater resistance to heat transfer through the insulation \cite{DOE2023}. The relative inefficiency of Texas building envelopes means that Texas building heating demand is particularly sensitive to cold weather. Texas buildings also are markedly reliant on electrical heating equipment for serving this heating demand. Sixty percent of residential buildings and 50\% of commercial buildings in Texas have primary space heating systems that are entirely powered by electricity \cite{wilson2021} (Figure 3a-b). These rates of electric heating adoption by the Texas residential and commercial building sectors are the 7th and 9th highest among U.S. states, respectively \cite{wilson2021}. Electric heating penetration has also been increasing over time, as in 1997 only 41\% of Texas residential buildings used electrical heating equipment as their primary space heating systems \cite{EIA2018}.

\begin{figure*}
\hfill
\begin{minipage}[b]{\textwidth}
\begin{subfigure}{0.5\textwidth}
\subcaption{}
\includegraphics[width=\textwidth]{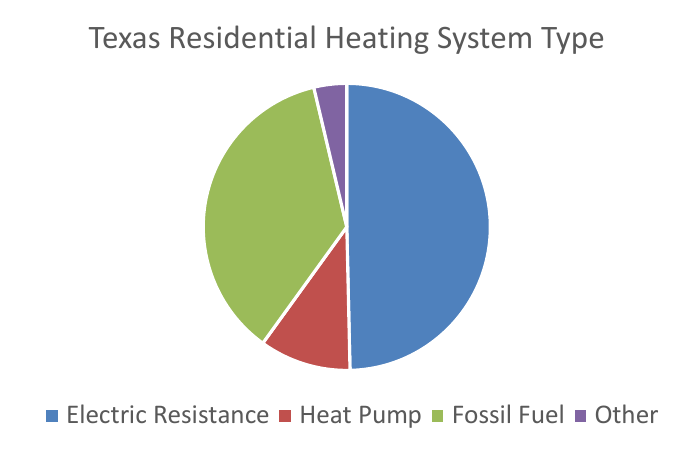}
\end{subfigure}
\begin{subfigure}[b]{0.5\textwidth}
\subcaption{}
\includegraphics[width=\textwidth]{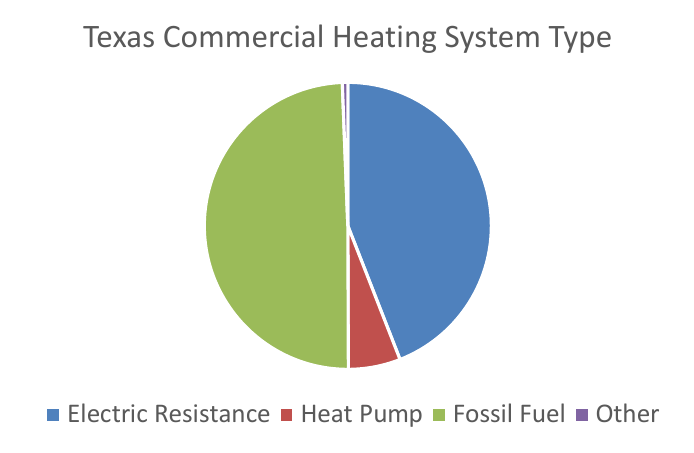}
\end{subfigure}\par
\begin{subfigure}[b]{\textwidth}
\subcaption{}
\centering
\includegraphics[width=\textwidth]{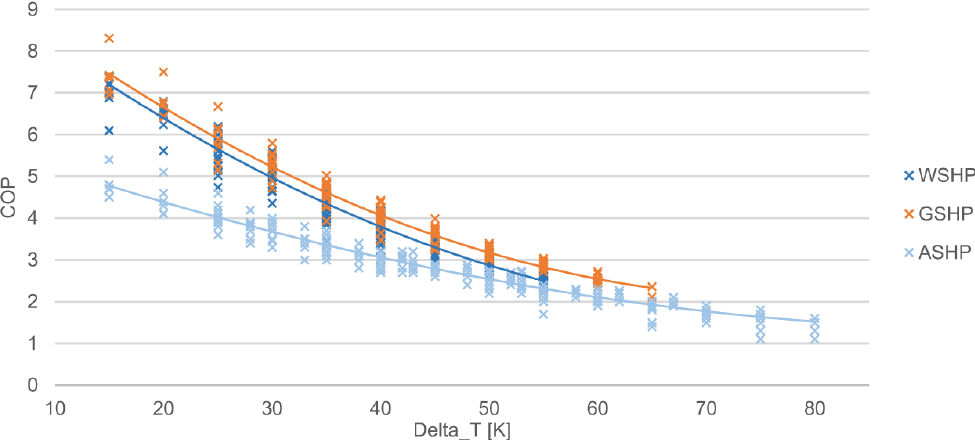}
\end{subfigure}
\caption{\textbf{a} and \textbf{b,} Most buildings in ERCOT are heated by electric resistance heating systems or heat pumps \cite{wilson2021}. \textbf{c,} The coefficient of performance (COP) for water-source heat pumps (WSHP), ground-source heat pumps (GSHP), and air-source heat pumps (ASHP) decreases with increasing temperature difference between the heat source and the heat sink. Figure from \cite{ruhnau2019}.}
\end{minipage}
\end{figure*}

\pagebreak

Additionally, the coefficient of performance (COP) for heat pumps diminishes as the temperature difference increases between the heat source and heat sink on either side of the building envelope (Figure 3c). COP in this case is defined as the amount of useful heat energy supplied to the building divided by the amount of electrical energy consumed by the heat pump. Thus, the amount of electricity required by a heat pump to meet an incremental increase in building heating load increases as outdoor temperatures drop. Furthermore, heat pump systems are often supplemented by energy-intensive backup electrical resistance heating systems (COP = 1) when the heat pump cannot fully meet building heating load. When outdoor temperatures fall below a certain threshold (e.g. 0°F (-17.8°C)), most heat pumps are designed to shut off and shift building heating load entirely on to these supplemental heating systems. The combination of poor envelope efficiency and high penetration of electrical heating in the Texas building stock could render the ERCOT power grid uniquely vulnerable to severe cold temperatures. However, meeting or updating codes for building envelopes and HVAC system efficiency could alter the sensitivity of ERCOT building electricity demand to severe temperatures and provide valuable grid reliability benefits.

In an effort to decarbonize building heating, government policies are seeking to hasten the adoption of electrical building heating equipment through technology mandates \cite{dichristopher2022, gruenwald2020, derrick2020} and incentive programs \cite{smedick2022, baldwin2022}. However, it is important to understand the performance of a highly electrified building stock under extreme winter weather conditions. The response of the ERCOT electrical system to Winter Storm Uri can be viewed as such a case study, as it represents an area with a high penetration of electrical building heating subjected to severe cold temperatures. This study will investigate ERCOT building electricity performance during Winter Storm Uri and assess the efficiency with which various building sector electrification and energy efficiency upgrades could have delivered power grid reliability benefits during the event.

\section{Methods}

Electricity demand in ERCOT is comprised of mostly stable industrial demand and weather-sensitive demand in the commercial and residential sectors (Figure 1a-b). Actual 2021 ERCOT sectoral electric load presented in Figure 1a-b was calculated by taking the fractional breakdown of served electric load for the three sectors in utility services areas that are open to retail competition \cite{ERCOT2021c} and applying those fractions to the total ERCOT served load \cite{ERCOT} to get system-wide sectoral served load. Residential, commercial, and industrial load fractions were defined as residential customers, business customers with peak load less than or equal to 700 kW and business customers with peak load greater than 700 kW, respectively \cite{ERCOT2021c}. In this analysis, actual served residential and commercial electric load were replaced with modeled residential and commercial electricity demand for different building stock scenarios. To assess the potential for building stock upgrades to reduce the need for load shed during Winter Storm Uri, modeled demand for these scenarios was compared to hourly served load \cite{ERCOT}, available generation \cite{ERCOT2023}, load shed \cite{ERCOT2023}, and estimated electric load without load shed \cite{ERCOT2023} data produced by ERCOT. Hourly served load is actual served load data produced by ERCOT for all of 2021. Available generation is the real-time generation on the system at that moment in time. Load shed is the ERCOT-requested load shed quantity. As utilities do not have fine control over the amount of load that they shed, the amount of load actually shed by utilities may have been higher than the load shed quantity requested by ERCOT. Estimated load without load shed is a back cast of load produced by ERCOT using the ERCOT Mid-Term Load Forecast process and actual weather data.

ERCOT residential and commercial energy demand were modeled using the National Renewable Energy Laboratory’s (NREL) ResStock \cite{wilson2017} and ComStock \cite{parker2023} models. Figure 4 presents a schematic of the ResStock modeling workflow. 

\begin{figure}[H]
\includegraphics[scale=0.55]{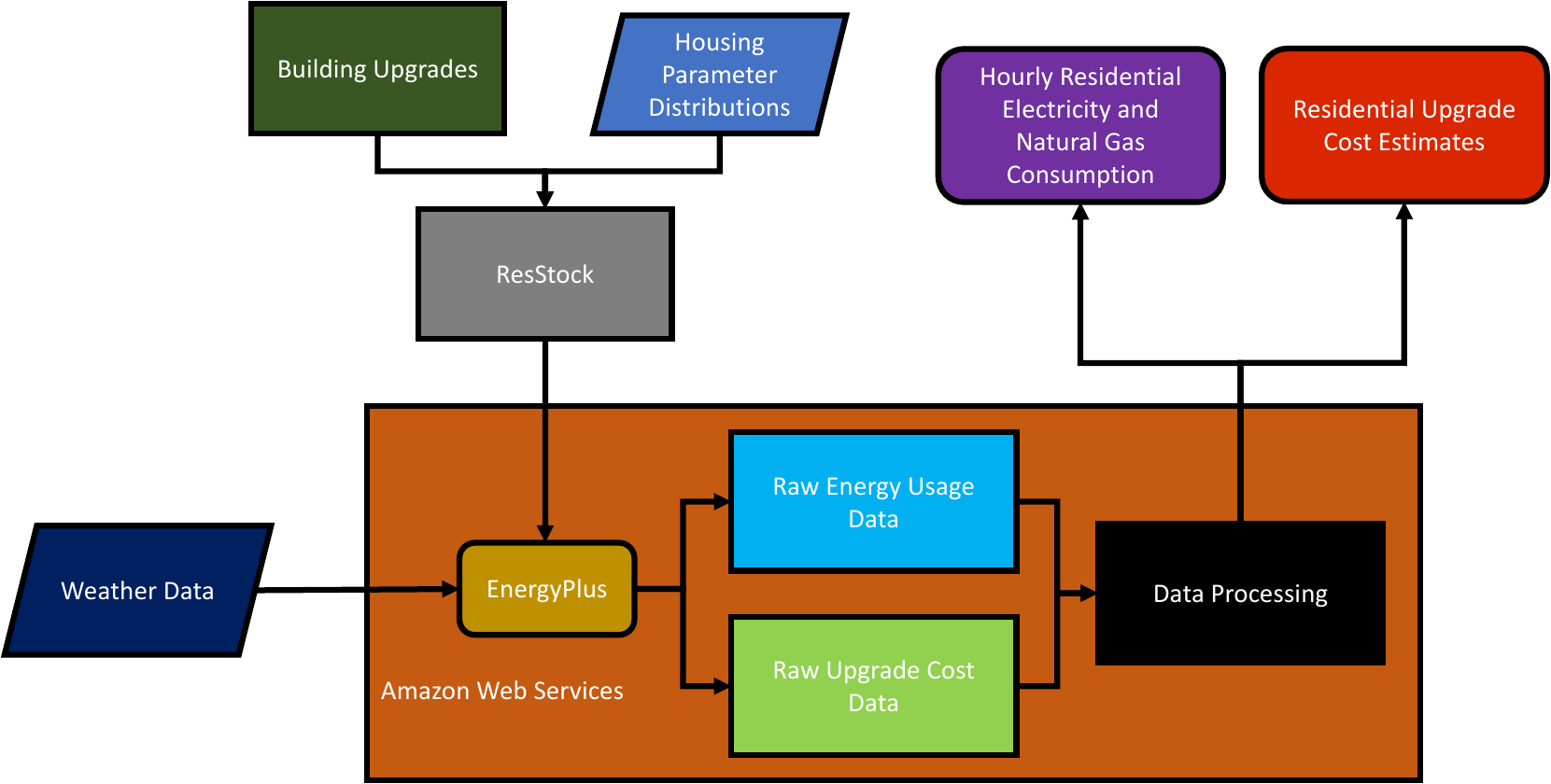}
\caption{Methodology flowchart for modeling residential sector energy demand with ResStock.}
\centering
\end{figure}

The core of the methodologies for both ResStock and ComStock is the EnergyPlus building energy simulation engine, developed by the U.S. Department of Energy \cite{crawley2001}. EnergyPlus is a building energy simulation software that considers building envelope characteristics and HVAC equipment parameters as it models building energy demand for heating and cooling according to thermal equations. EnergyPlus also models energy demand for other loads such as people, computers, and lighting. ResStock and ComStock both generate representative building stock samples based on spatially resolved commercial and public data sources such as the Residential Energy Consumption Survey (RECS) and the Commercial Buildings Energy Consumption Survey (CBECS) administered by the U.S. Energy Information Administration (EIA) \cite{wilson2021}. The building stock characteristics can then be altered to reflect different building upgrades e.g., installing a certain level of insulation or a specific air-source heat pump (ASHP) model in every building. Each building from these building stock samples is then modeled with EnergyPlus according to building characteristics and local weather data. In both the ResStock and ComStock models, heating and cooling systems are auto sized based on building-specific parameters. If the primary building heating system is a heat pump and heating load exceeds the heat pump capacity, then a supplemental heating system activates to meet the remaining heating load. Supplemental heating systems in the baseline building stock are comprised of both fossil-fueled and electric resistance heating systems. All building stock scenarios in this study that include a ASHP retrofit are coupled with supplemental heating systems that are entirely reliant on electric resistance heating. All ASHPs installed via retrofit in this study shut off if outdoor temperatures drop below 0°F (-17.8°C). If the combined primary and supplementary heating systems cannot fully meet building heating load, indoor temperatures will drop below building thermostat setpoints. EnergyPlus simulation outputs include high resolution electricity, natural gas, and other fuel consumption for the entire building and for numerous subloads (e.g. heating, cooling, lighting). Because each building is modeled separately by EnergyPlus, the modeling process can be parallelized, allowing for many building simulations to be run at the same time. Small-scale simulations (less buildings) can be modeled on a personal computer, whereas larger scale simulations need to be deployed on a high-performance computing (HPC) asset. These building stock samples are designed to be statistically representative of the entire sector, thus the aggregate modeled energy demand for the building stock samples can be linearly scaled to represent energy demand for the entire sector.

\subsection{ERCOT population-weighted weather metrics}
For this analysis, we also consider an ERCOT population-weighted degree day (DD) metric that represents aggregate ERCOT building heating or cooling demand estimated with spatially-resolved weather data \cite{skiles2023}. A heating degree day (HDD) or cooling degree day (CDD) compares the outdoor temperature for a location to a standard temperature (usually near a typical building thermostat setpoint, such as 18.5°C or 65.3°F) as a measure of heating and cooling load, respectively \cite{EIA2021a}. In this study, ERCOT-wide population-weighted DD metrics were calculated by summing the DD from the largest metropolitan area in each of the eight ERCOT weather zones (Figure 5 and Table 1) and weighting the DD values by the population in each weather zone \cite{skiles2023}. Weather data was sourced from EnergyPlus weather (EPW) files \cite{smith2021} and population data was sourced from the United States Census Bureau \cite{uscensus}. 

We also developed an ERCOT population-weighted mean temperature metric. This metric is defined as the ambient temperature that corresponds to the ERCOT population-weighted HDD.

\begin{equation}
Hourly\; ERCOT\; mean\; temp\; = setpoint\; temp\; -\; hourly\; ERCOT\; HDD\; \times 24\;
\end{equation}

The hourly ERCOT population-weighted mean temperature metric represents the ambient temperature that, if applied at each of the eight weather stations, would produce an equivalent ERCOT population-weighted HDD value as was calculated with the eight independent temperature readings from that hour. It should be noted that outside of the week of Winter Storm Uri, weather stations experienced CDD (not HDD). The ERCOT population-weighted mean temperature metric does not capture the cooling load experienced at those weather stations during those hours.

\begin{figure}[H]
\includegraphics[scale=0.2]{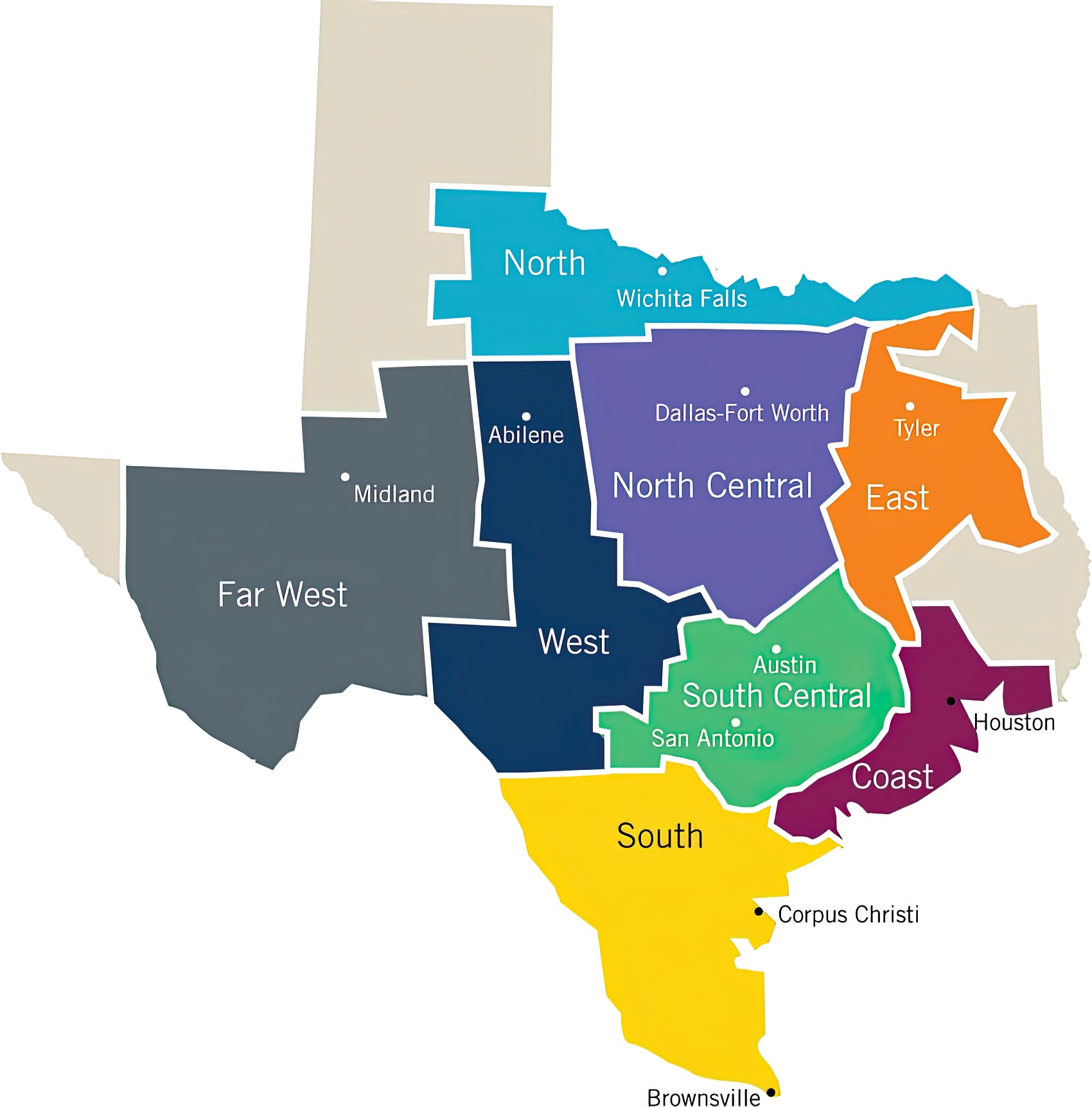}
\caption{Map of the eight ERCOT weather zones \cite{ERCOTWeatherZone}.}
\centering
\end{figure}

\newcolumntype{P}[1]{>{\raggedright\arraybackslash}p{#1}}
\newcolumntype{M}[1]{>{\centering\arraybackslash}m{#1}}
\begin{table}
  \centering
  \begin{tabular}{|M{2.3cm}|M{2.3cm}|M{1.7cm}|M{2.7cm}|M{2.7cm}|} 
    \hline
    Weather Zone & City & Weather Station ID & 2021 Average Summer Hourly CDD & 2021 Average Winter Hourly HDD \\ \hline\hline
    Coast & Houston & KIAH & 0.42 & 0.23 \\ \hline
    East & Tyler & KTYR & 0.38 & 0.36 \\ \hline
    Far West & Midland & KMAF & 0.32 & 0.45 \\ \hline
    North & Wichita Falls & KSPS & 0.36 & 0.49 \\ \hline
    North Central & DFW & KDFW & 0.42 & 0.38 \\ \hline
    South & Corpus Christi & KCRP & 0.41 & 0.19 \\ \hline
    South Central & Austin & KAUS & 0.36 & 0.29 \\ \hline
    West & Abilene & KABI & 0.38 & 0.41 \\ \hline
  \end{tabular}
  \caption{2021 average seasonal degree day information for the weather stations used for each of the eight ERCOT weather zones.}\label{tab1}
\end{table}

\pagebreak

\subsection{ResStock}
Researchers can deploy ResStock with custom upgrade scenarios and weather files. For this study, large-scale ResStock simulations were deployed on Amazon Web Services (AWS), as shown in Figure 4. Because building electricity demand for space conditioning is driven by weather patterns, it was important that each EnergyPlus building model used high-resolution, spatially resolved weather files. In this study, this was achieved by utilizing weather files from the largest metropolitan area in each of the eight ERCOT weather zones. For each EnergyPlus simulation, the county in which the building was located was mapped to the corresponding ERCOT weather zone and the weather file for the largest metropolitan area in that weather zone. This allowed the ResStock model to capture the impact of local weather on building energy demand. In this study, 2021 actual meteorological year (AMY 2021) data was used \cite{smith2021}. This allowed for the estimation of ERCOT residential energy demand using actual weather data from Winter Storm Uri in February 2021. Modeled residential electricity demand was also apportioned to space heating demand, space cooling demand, and other (non-space heating and cooling) demand based on sub-loads. 

An underlying assumption of the ResStock model is that energy demand for the building stock sample is statistically representative of energy demand for all buildings in the residential sector. Thus, the larger the size of the building sample that is modeled, the more accurate the estimation of sector-wide energy demand. Therefore, choosing a building stock sample size is a matter of balancing model accuracy with increasing computing costs for larger samples. To determine the building stock sample size necessary to deliver the desired accuracy, a convergence study was conducted. A range of building sample sizes with baseline building characteristics were modeled with ResStock and average building electricity consumption for each sample was calculated (Figure 6). It was determined that further increases in building sample size beyond \(\sim \)5000 buildings did not materially alter the average modeled baseline building electricity consumption for the entire 2021 calendar year or the month of February 2021 (the period of interest). Thus, the sample size for ResStock simulations in this study was set at \(\sim \)5000 buildings. The cumulative energy demand for the \(\sim \)5000 building simulations was then linearly scaled to represent the \(\sim \)9.3 million residential buildings in ERCOT, as defined by the ResStock model.

\begin{figure}[H]
\includegraphics[scale=0.75]{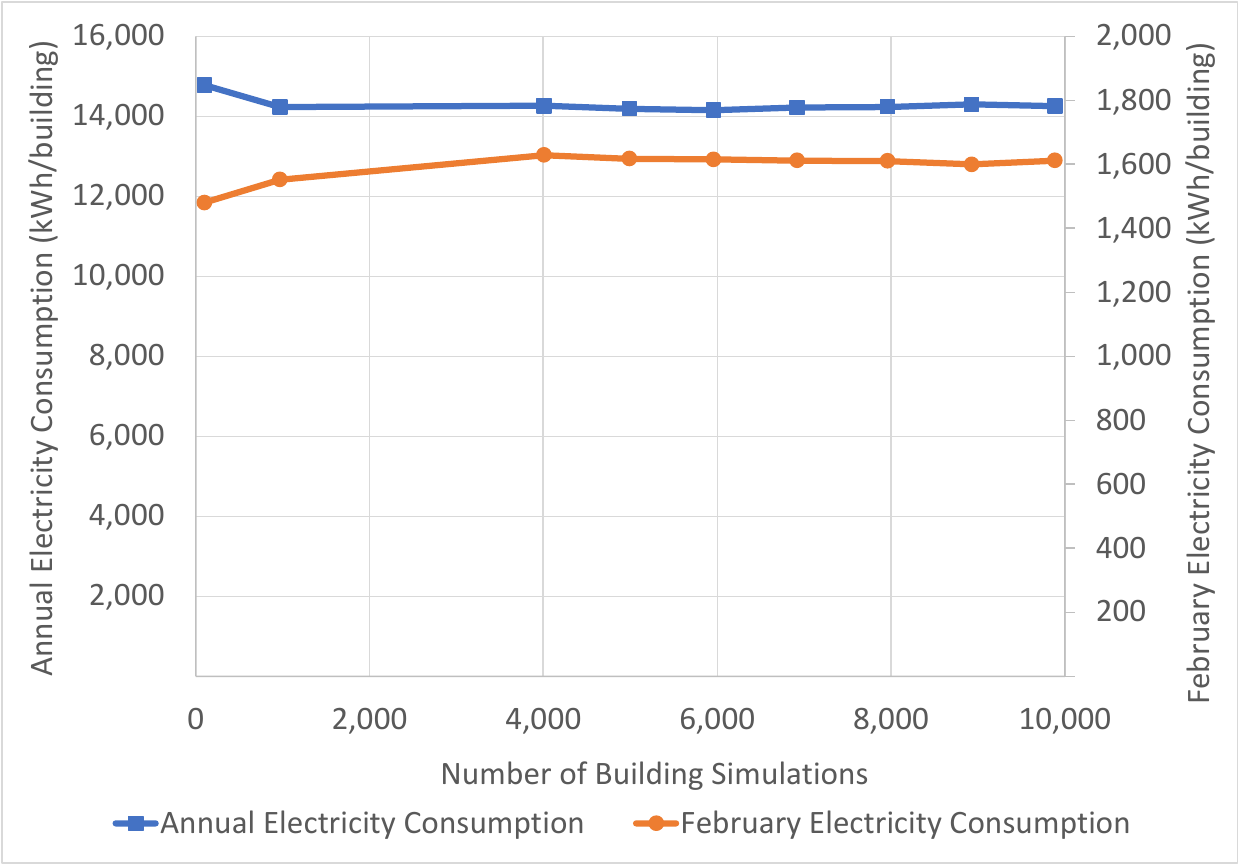}
\caption{Sensitivity testing of ResStock building stock sample size indicated that average modeled baseline building electricity consumption converged at a building stock sample size of around 5000 buildings.}
\centering
\end{figure}

ResStock was used to model ERCOT residential sector electricity demand for the baseline building stock as well as for the baseline building stock with all buildings modified by three upgrade packages: an efficiency package, an electrification package, and a combined efficiency + electrification package. These ResStock upgrade packages do not include demand response measures and are comprised of the static building retrofits shown in Table 2. 

\pagebreak

\begin{table}
\centering
  \begin{tabular}{|c|c|c|} 
    \hline
    Upgrade Package & Parameter & Upgrade \\ \hline\hline
    \multirow{3}{*}{Efficiency}
         & \multicolumn{1}{|l|}{Ceiling Insulation} & \multicolumn{1}{|l|}{R-38} \\\cline{2-3}
         & \multicolumn{1}{|l|}{Infiltration} & \multicolumn{1}{|l|}{7 ACH50} \\\cline{2-3}
         & \multicolumn{1}{|l|}{Ducts} & \multicolumn{1}{|l|}{7.5\% Leakage and R-8 Insulation} \\ \hline
    \multirow{1}{*}{Electrification}
         & \multicolumn{1}{|l|}{Heating/Cooling System} & \multicolumn{1}{|l|}{ASHP, SEER 18, 9.3 HSPF} \\ \hline
    \multirow{4}{*}{Efficiency + Electrification}
         & \multicolumn{1}{|l|}{Ceiling Insulation} & \multicolumn{1}{|l|}{R-38} \\\cline{2-3}
         & \multicolumn{1}{|l|}{Infiltration} & \multicolumn{1}{|l|}{7 ACH50} \\\cline{2-3}
         & \multicolumn{1}{|l|}{Ducts} & \multicolumn{1}{|l|}{7.5\% Leakage and R-8 Insulation} \\\cline{2-3}
         & \multicolumn{1}{|l|}{Heating/Cooling System} & \multicolumn{1}{|l|}{ASHP, SEER 18, 9.3 HSPF} \\\hline
  \end{tabular}
  \caption{ResStock was used to model ERCOT residential sector electricity demand for building stock with three upgrade package scenarios: an efficiency package, an electrification package, and a combined efficiency + electrification package.}\label{tab1}
\end{table}

For each residential building, if the ceiling insulation, infiltration, or ducts parameters were less efficient than the upgraded condition, the parameter was upgraded. If the parameter in the original building stock sample was already as efficient or more efficient than the upgraded condition, the parameter was left as is. Similarly, for the electrification upgrade, if the heating system in the original building stock sample was less efficient than an ASHP, SEER 18, 9.3 HSPF or the equipment was fossil-fueled, the heating and cooling systems were upgraded to an ASHP, SEER 18, 9.3 HSPF. If the heating system in the original building stock sample was electric and had an efficiency that was equal to or greater than the ASHP, SEER 18, 9.3 HSPF efficiency, then the electrification upgrade was not applied. 

\subsection{Comstock}
In this study, ERCOT’s commercial sector electricity demand was sourced from a database of results for ComStock simulations previously run by NREL personnel on NREL’s Eagle HPC asset \cite{wilson2021}. Modeled commercial electricity demand was apportioned to space heating demand, space cooling demand, and other (non-space heating and cooling) demand based on sub-loads. These simulations were run with AMY 2018 weather files for each county. For counties in which no weather data was available, AMY 2018 weather data from the closest weather station in the same climate zone was used \cite{parker2023}. The ComStock model has only been developed to model commercial building types (e.g. hospital, large office, primary school) that comprise 64\% of U.S. commercial building floor space. It was assumed that this subsection of commercial building floorspace was representative of the entire commercial building sector, and total modeled energy demand for this segment of the ERCOT commercial building sector was linearly scaled to represent all commercial building floor space in ERCOT. 

To adjust modeled AMY 2018 electricity demand to account for AMY 2021 weather, second order polynomial regressions were developed for each weekday hour and each weekend hour between modeled hourly heating and cooling electricity demand and hourly AMY 2018 ERCOT population-weighted degree-days. This resulted in 48 polynomial regressions each for heating load and cooling load. Second-order polynomial regressions were chosen due to a previous analysis that identified a second-order polynomial relationship between per capita ERCOT electricity demand and ERCOT population-weighted degree days \cite{skiles2023}. Then for each hour in 2021, the hourly ERCOT population-weighted degree-days calculated using AMY 2021 weather data was plugged into the corresponding AMY 2018 hourly regression to estimate commercial sector heating and cooling electricity demand for that hour in 2021. Because other (non-space heating and cooling) electricity demand is not weather-sensitive, AMY 2021 other electricity demand for each hour of the weekday and weekend was taken as the annual average modeled demand for the corresponding hour in the modeled AMY 2018 demand dataset. Methods for adjusting AMY 2018 electricity demand data for AMY 2021 weather are outlined in Table 3.

\begin{table}[H]
\centering
  \begin{tabular}{|c|c|c|c|} 
    \hline
    Hour of the Week & Independent Variable & Dependent Variable & Modeling Strategy \\ \hline\hline
    \multirow{3}{*}{Weekday Hour}
         & \multicolumn{1}{|l|}{ERCOT population-weighted HDD} & \multicolumn{1}{|l|}{Space Heating Demand} & \multicolumn{1}{|l|}{2nd Order Polynomial}\\\cline{2-4}
         & \multicolumn{1}{|l|}{ERCOT population-weighted CDD} & \multicolumn{1}{|l|}{Space Cooling Demand} & \multicolumn{1}{|l|}{2nd Order Polynomial}\\\cline{2-4}
         & \multicolumn{1}{|c|}{-} & \multicolumn{1}{|l|}{Other Demand} & \multicolumn{1}{|l|}{Annual Average} \\ \hline
    \multirow{3}{*}{Weekend Hour}
         & \multicolumn{1}{|l|}{ERCOT population-weighted HDD} & \multicolumn{1}{|l|}{Space Heating Demand} & \multicolumn{1}{|l|}{2nd Order Polynomial}\\\cline{2-4}
         & \multicolumn{1}{|l|}{ERCOT population-weighted CDD} & \multicolumn{1}{|l|}{Space Cooling Demand} & \multicolumn{1}{|l|}{2nd Order Polynomial}\\\cline{2-4}
         & \multicolumn{1}{|c|}{-} & \multicolumn{1}{|l|}{Other Demand} & \multicolumn{1}{|l|}{Annual Average} \\ \hline
  \end{tabular}
  \caption{ERCOT hourly commercial electricity demand for specific end-uses modeled with ComStock using AMY 2018 data was adjusted to account for AMY 2021 weather by applying averages of AMY 2018 data or using regressions of population-weighted HDD.}\label{tab1}
\end{table}

ComStock was used to model commercial sector electricity demand for the baseline building stock as well as for the baseline building stock upgraded where applicable with an efficiency package, an electrification package, and a combined efficiency + electrification package. These ComStock upgrade packages do not include demand response measures and are comprised of the static building retrofits presented in Table 4.

\begin{table} [H]
\newcolumntype{M}[1]{>{\centering\arraybackslash}m{#1}}
\newcommand{\minitab}[2][l]{\begin{tabular}{#1}#2\end{tabular}}
\newcolumntype{L}[1]{>{\raggedright\let\newline\\\arraybackslash\hspace{0pt}}m{#1}}
\centering
  \begin{tabular}{|M{3cm}|L{3.2cm}|L{8cm}|}
    \hline
    \multicolumn{1}{|>{\centering\arraybackslash}m{3cm}|}{Upgrade Package}
    & \multicolumn{1}{|>{\centering\arraybackslash}m{3.2cm}|}{Parameter}
    & \multicolumn{1}{|>{\centering\arraybackslash}m{8cm}|}{Upgrade}\\ \hline\hline
    \multirow{5}*{Efficiency}
         & Wall Insulation & Add rigid insulation under exterior cladding outside structural elements to meet climate zone-specific target R-values specified by the Advanced Energy Design Guidelines (AEDG). \\\cline{2-3}
         & Roof Insulation & Increase roof insulation to align with climate zone-specific AEDG target R-values. \\\cline{2-3}
         & Windows & Replace existing windows with those that align with climate zone-specific AEDG properties. \\ \hline
    \multirow{6}*{Electrification}
         & Lighting & Replace all interior lighting with LED lighting. \\\cline{2-3}
         & Heat Pump Rooftop Units (HP-RTU) & Replace gas and electric resistance RTUs with variable speed, high efficiency (>17 IEER) HP-RTUs. \\\cline{2-3}
         & Heat Pump Boiler (HP-boiler) & Replace natural gas boilers for space heating with air-source HP-boilers. \\ \hline
    \multirow{10}*{\minitab[c]{Efficiency + \\ Electrification}}
         & Wall Insulation & Add rigid insulation under exterior cladding outside structural elements to meet climate zone-specific target R-values specified by the Advanced Energy Design Guidelines (AEDG). \\\cline{2-3}
         & Roof Insulation & Increase roof insulation to align with climate zone-specific AEDG target R-values. \\\cline{2-3}
         & Windows & Replace existing windows with those that align with climate zone-specific AEDG properties. \\\cline{2-3}
         & Lighting & Replace all interior lighting with LED lighting. \\\cline{2-3}
         & Heat Pump Rooftop Units (HP-RTU) & Replace gas and electric resistance RTUs with variable speed, high efficiency (>17 IEER) HP-RTUs. \\\cline{2-3}
         & Heat Pump Boiler (HP-boiler) & Replace natural gas boilers for space heating with air-source HP-boilers. \\ \hline
  \end{tabular}
  \caption{ComStock was used to model ERCOT commercial sector electricity demand for building stock with three upgrade package scenarios: an efficiency package, an electrification package, and a combined efficiency + electrification package \cite{NREL2023a}.}\label{tab1}
\end{table}

\subsection{Adjusting for modeling bias}
In this analysis, actual served residential and commercial electricity demand was replaced with electricity demand for different building stock scenarios as modeled with ResStock and ComStock. To adjust for bias in the ResStock modeling, the percent difference between ResStock-modeled residential baseline electricity consumption and actual served residential electricity load for each month in 2021 was applied to the hourly ResStock-modeled electricity demand for the corresponding month. This percent adjustment to modeled residential baseline electricity demand ensured that modeled residential baseline electricity consumption equaled actual served residential electricity load for each month. An exception was made for February 2021, where the percent difference between January 2021 modeled baseline electricity consumption and January 2021 actual served electricity load was applied to February 2021 hourly modeled demand data. This exception was made because February 2021 actual served residential electricity load measured by ERCOT is an underestimate of demand, as ERCOT was forced to administer load shed to balance supply and demand. The percent differences between modeled monthly residential baseline electricity consumption and actual monthly served residential electricity load were also applied to modeled residential electricity demand for each upgrade scenario. Actual monthly served commercial electricity load data was used to apply this same percent adjustment methodology to modeled commercial electricity demand data. The percent difference between ComStock-modeled AMY 2021 baseline electricity consumption and actual served commercial electricity load for each month in 2021 was applied to the hourly ComStock-modeled AMY 2021 baseline electricity demand. Due to load shed during February 2021, the percent difference between January 2021 modeled baseline commercial consumption and January 2021 actual served commercial electricity load was applied to modeled February 2021 hourly baseline electricity demand. The percent adjustment for the modeled baseline commercial demand was also applied to ComStock-modeled electricity demand for each of the building upgrade scenarios. These adjusted residential and commercial demand profiles were used as modeled demand in this analysis.

One limitation with this analysis is that the modeled residential and commercial electricity demand produced with ResStock and ComStock appear to be more sensitive to ambient temperatures and occupancy schedules than actual building electricity demand. While the monthly percent adjustments ensured that modeled baseline monthly total electricity consumption equaled actual total served electricity load for each month (except February), the modeled hourly baseline demand peaks tended to be higher and the troughs tended to be lower than actual hourly served load. For example, modeled baseline scenario summer coincident peak demand occurred on Tuesday August 24th at 81.3 GW, while ERCOT measured a maximum summer coincident peak load of 73.7 GW.

\section{Results and Discussion}

ERCOT 2021 electricity end-use demand is presented in Figure 7. Outside of February, demand in Figure 7 is comprised of actual served industrial load and actual served residential and commercial load broken down into end-uses based on fractional end-use data from ResStock and ComStock modeling of baseline building stock. During February, demand is comprised of actual served industrial load and modeled baseline residential and commercial demand.

\begin{figure*}
\hfill
\begin{minipage}[b]{\textwidth}
\begin{subfigure}{\textwidth}
\subcaption{}
\includegraphics[scale=0.4]{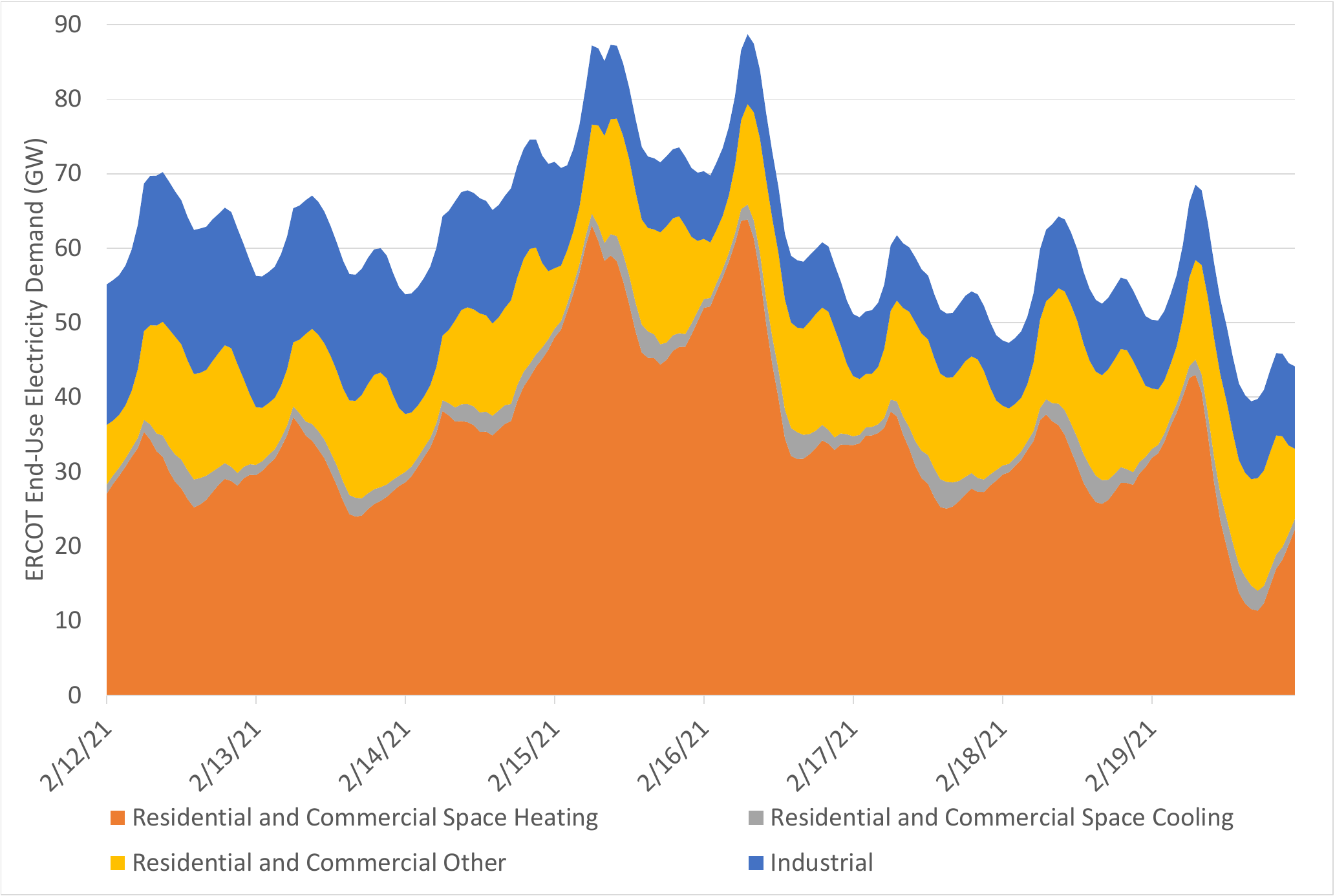}
\centering
\end{subfigure}\par
\begin{subfigure}{\textwidth}
\subcaption{}
\includegraphics[scale=0.4]{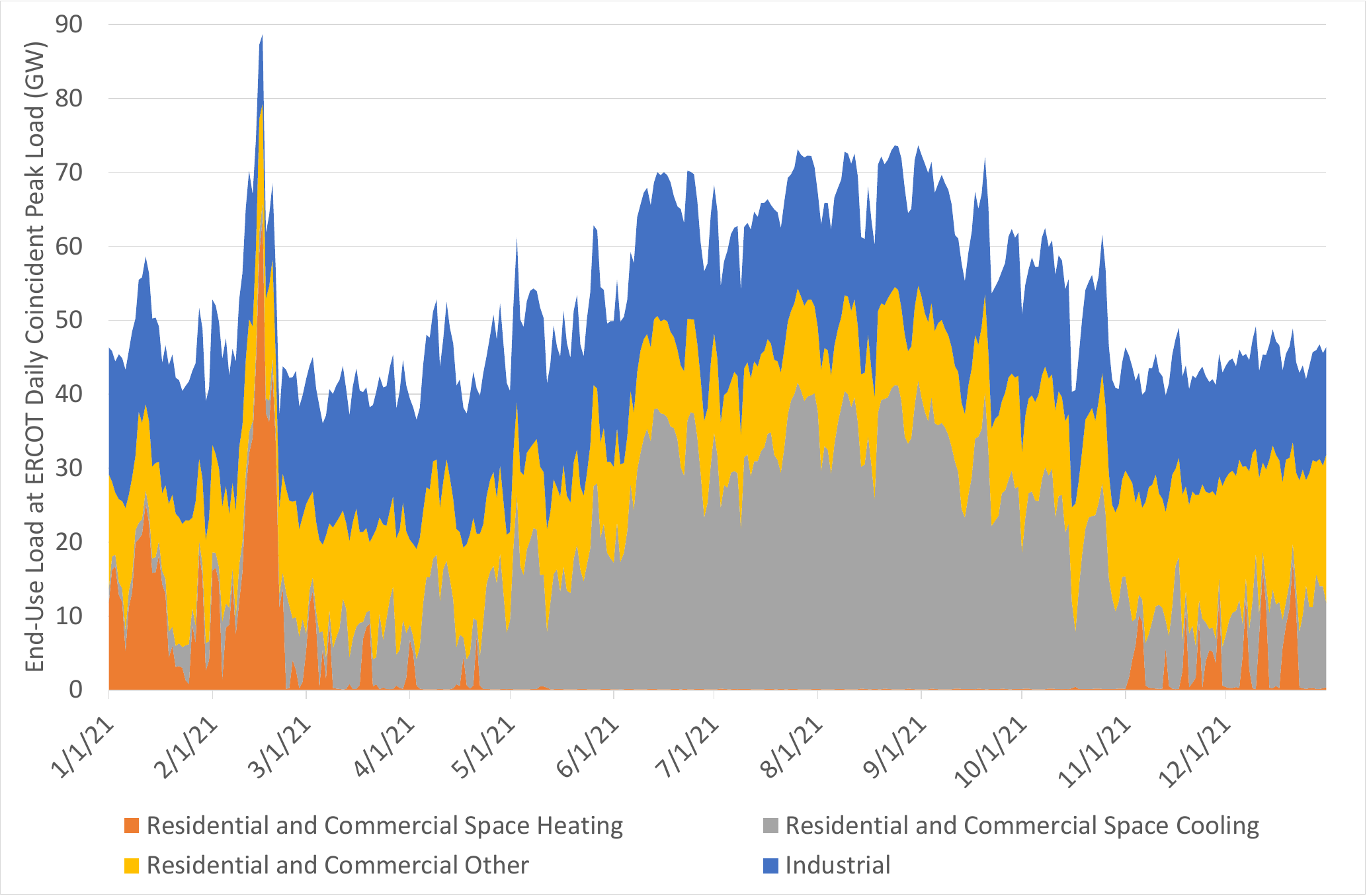}
\centering
\end{subfigure}
\end{minipage}
\quad
\begin{minipage}[b]{\textwidth}
\caption{\textbf{a,} Residential and commercial building space heating was the primary driver of peak electricity demand during Winter Storm Uri. \textbf{b,} Building space heating and cooling loads were the drivers of all peak demand events in 2021.}\label{key}
\centering
\end{minipage}
\end{figure*}

\pagebreak{}
Actual served ERCOT industrial load was fairly uniform and varied between 13.6 GW and 22.1 GW during daily coincident peak demand hours outside of February. During Winter Storm Uri, actual served industrial load was more variable because power flow to industrial consumers was curtailed so available electricity could be allocated to more critical building loads. Similarly, ERCOT residential and commercial sectors consistently required between 8.4 GW and 21.8 GW of electricity for other loads (non-space heating and cooling loads) during daily coincident peak hours throughout the year. Conversely, residential and commercial building cooling systems required anywhere from 1.6 GW up to 41.8 GW during daily peak demand hours depending on the weather. During the summer, residential and commercial building cooling accounted for up to 57\% of daily coincident peak demand. Residential and commercial building heating systems needed anywhere from 0.1 GW to 63.8 GW depending on ambient temperatures. During Winter Storm Uri, residential and commercial building heating systems were responsible for up to 72\% of ERCOT electricity demand. This spike in demand during Winter Storm Uri was a consequence of the Texas building stock that had relatively inefficient envelopes and high penetration of lower efficiency electrical heating systems. As demonstrated in Figure 3c, the COP of a typical heat pump diminishes as the temperature difference between the hot and cold reservoirs increases. At very cold temperatures, buildings with heat pump systems increasingly rely on energy-intensive backup electric resistance heating systems. Much of the space heating demand modeled during Winter Storm Uri can be attributed to these equipment factors. These demand trends demonstrate the importance of building heating and cooling loads and highlight the need to effectively manage this portion of demand.

Hourly ERCOT demand during Winter Storm Uri for the baseline building stock scenario is presented in Figure 8 along with available generation \cite{ERCOT2023} and ERCOT-estimated demand. ERCOT-estimated demand is defined as estimated load without load shed for hours in which ERCOT requested load shed from utilities. For hours in which ERCOT did not request utilities to shed load, ERCOT-estimated demand is defined as actual served load. Figure 8 also includes hourly population-weighted mean temperature.

\begin{figure}[H]
\includegraphics[scale=0.45]{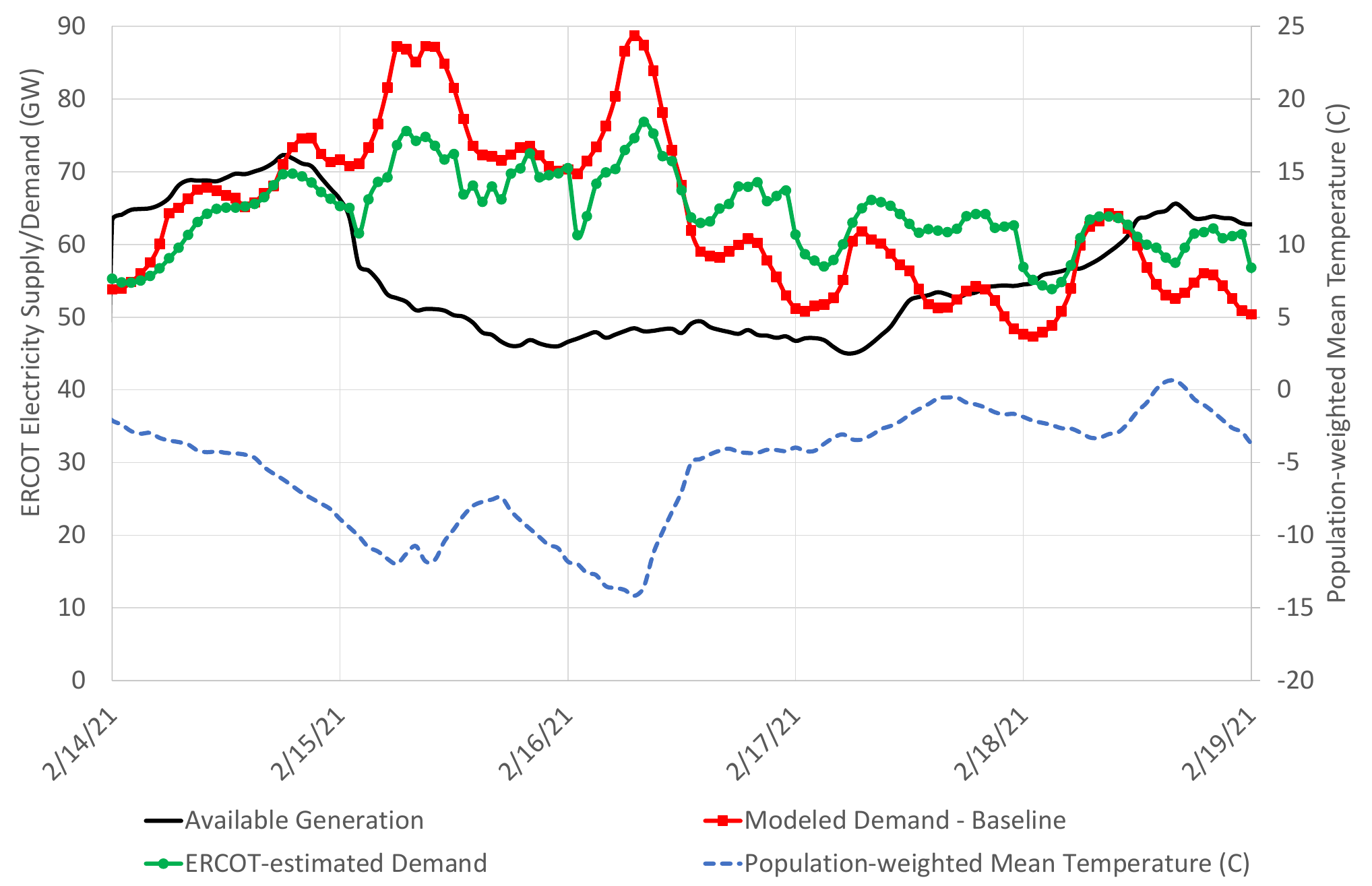}
\caption{Modeled hourly ERCOT demand for the baseline scenario had higher peaks, but otherwise closely tracked with ERCOT-estimated demand. Both demand profile estimates were inversely correlated with population-weighted mean temperature and significantly exceeded available generation starting early in the morning of February 15, 2021, resulting in an electricity shortfall.}
\centering
\end{figure}

Electricity demand for the baseline scenario (red line with squares), closely tracked with ERCOT-estimated demand (green line with circles), lending validity to the ComStock and ResStock modeling framework. Baseline scenario electricity demand did peak at 87.3 GW and 88.7 GW during the coldest hours of the week on February 15\textsuperscript{th} and 16\textsuperscript{th}, which was higher than the corresponding daily peaks in demand estimated by ERCOT (75.6 GW and 76.8 GW). However, previous research has estimated daily peak demand was 82.2 GW on February 15\textsuperscript{th} and 81.9 GW on February 16\textsuperscript{th} \cite{lee2022b}, which is more comparable to our estimated demand for the baseline scenario. During the middle of the week, when ERCOT generators could not keep pace with demand, available generation (black line) dropped below modeled baseline demand and ERCOT-estimated demand. The area between demand and available generation (for hours in which demand exceeds available generation) represents the electricity shortfall during the event. Both the modeled baseline demand and the ERCOT-estimated demand were inversely correlated with population-weighted mean temperature (blue dashed line) and peaked on February 15\textsuperscript{th} and 16\textsuperscript{th} when population-weighted mean temperature was at its minima. Hourly ERCOT electricity demand during Winter Storm Uri for our three alternative building stock scenarios (efficiency, electrification, and efficiency + electrification) is presented in Figure 9 along with available generation \cite{ERCOT2023}.

\begin{figure}[H]
\includegraphics[scale=0.3]{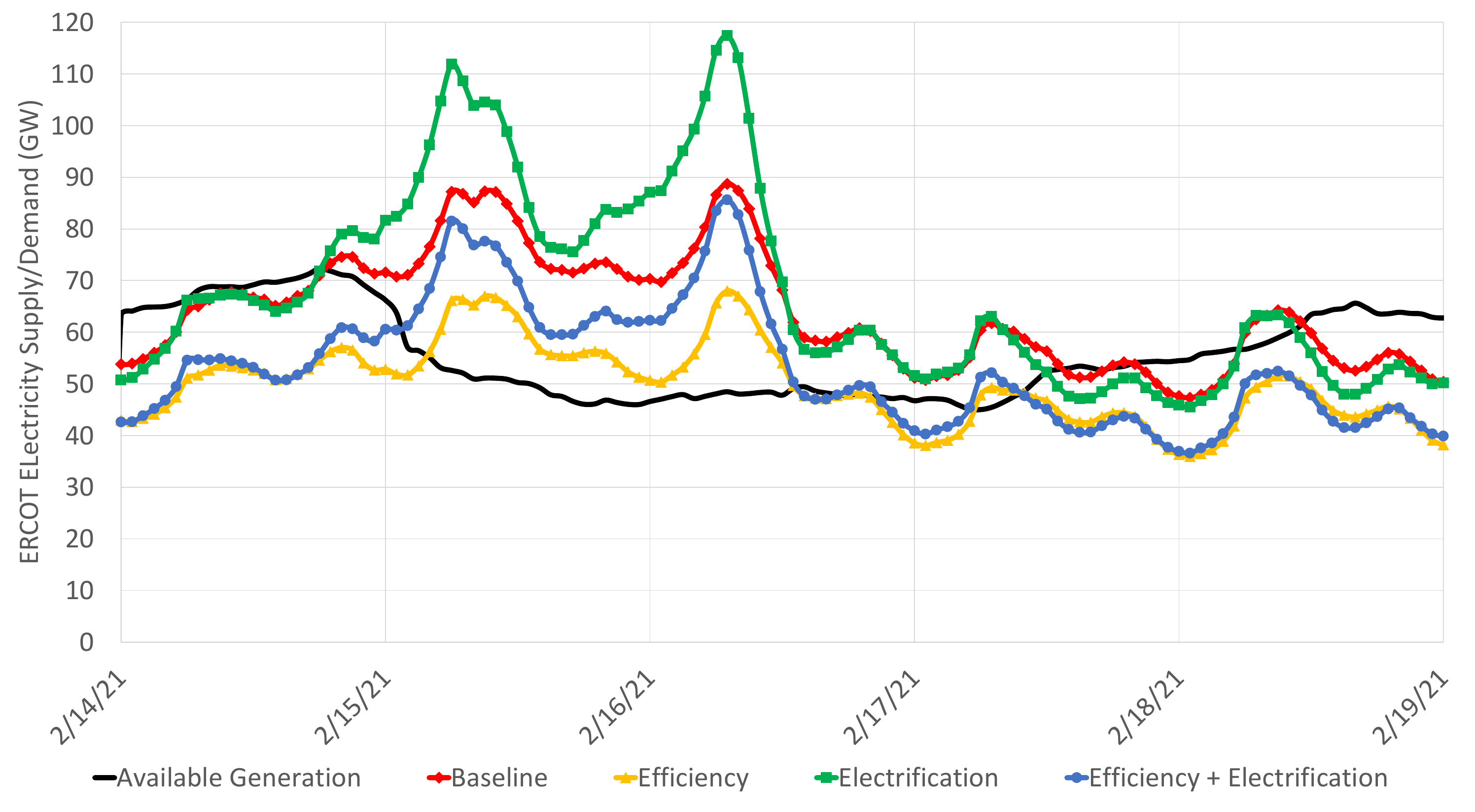}
\caption{Hourly ERCOT electricity demand for different upgrade scenarios indicates that the electrification upgrade package would have increased peak electricity demand during Winter Storm Uri, while the efficiency and the combined efficiency + electrification upgrade packages would have reduced overall consumption and peak demand during Winter Storm Uri.}
\centering
\end{figure}

The area between available generation (black line) and modeled electricity demand for the other building stock scenarios represents the electricity shortfall (from 2/15/21 through 2/17/21) that would have occurred for each building stock scenario. For the scenario in which all commercial and residential buildings received an electrification upgrade (green line with squares), ERCOT electricity demand would have hit two distinct peaks of 111.8 GW and 117.5 GW. These peaks would have been 28\% and 32\% higher than baseline scenario (red line with diamonds) demand peaks for the corresponding days (87.3 GW and 88.7 GW). However, electricity demand for the electrification scenario would only have greatly outpaced baseline demand on February 15\textsuperscript{th} and 16\textsuperscript{th} when hourly ERCOT population-weighted mean temperature reached lows of -12.0°C (10.5°F) and -14.2°C (6.5°F), respectively. Electrification scenario demand would have been comparable to baseline scenario demand the rest of the week. The divergence in the efficiencies of ASHPs and baseline heating equipment in the middle of the week can be attributed to the dramatic drop in ambient temperature. As outdoor temperatures drop, ASHP efficiency decreases, and these systems must increasingly rely on energy-intensive backup electric resistance heating. The impact of these technological limitations was evident as the ERCOT population-weighted mean temperature dropped from 1.2°C (34.1°F) at 4:00 PM on February 13\textsuperscript{th} to -12.0°C (10.5°F) at 6:00 AM on February 15\textsuperscript{th}, and ERCOT electricity demand for the electrification scenario more than doubled. 

These phenomena are a testament to the impressive efficiency of modern ASHP equipment at moderate ambient temperatures. Even after absorbing the heat generation load typically exerted on the natural gas system, system-wide ASHP space heating still would have reduced overall electricity demand relative to the baseline building stock at ambient temperatures as low as -3.3°C (26.0°F). However, as ERCOT population-weighted mean temperature dropped below -3.3°C (26.0°F), diminishing ASHP efficiency and increasing demand from electrical resistance heating would have driven system electricity demand well above baseline demand. This would have resulted in significantly more load shed during Winter Storm Uri. It should be noted however, that the avoided direct consumption of natural gas by building heating systems would have freed up fuel in the strained natural gas delivery system. This could have resulted in additional power generation, as a substantial portion of outages at power generation facilities were attributed to fuel limitations \cite{king2021}. These trends for the electrification scenario are especially relevant as global institutions continue to incentivize heat pump adoption as a decarbonization tactic. Non-cold weather heat pumps like those considered in this study can contribute to dramatic spikes in electricity demand when installed in poorly insulated buildings and subjected to severe cold temperatures. Resource planners should consider incentivizing higher efficiency cold-weather heat pumps combined with envelope efficiency retrofits as they seek to electrify building stock that experiences severe cold temperatures.

Electricity demand for the efficiency upgrade scenario (yellow line with triangles) would have been lower than baseline electricity demand during every hour of Winter Storm Uri. The efficiency upgrade package would have reduced total ERCOT electricity consumption by 23\% from February 15\textsuperscript{th} through February 17\textsuperscript{th}, the days in which ERCOT-requested load shed. The efficiency upgrade package would have lowered peak demand for the baseline building stock (88.7 GW) down to 68.0 GW. Thus, the efficiency upgrade package would have provided the same grid service during Winter Storm Uri as 20.7 GW of additional dispatchable power generation. The efficiency upgrade package would have lowered ERCOT demand below available generation for most of the week, thus averting the need for much of the load shed administered by ERCOT. Electricity demand for the efficiency + electrification upgrade scenario (blue line with circles) also would have been lower than baseline electricity demand during every hour of the event. This means that the efficiency + electrification upgrade package would have reduced the need for load shed. Additionally, natural gas that would have been saved in buildings due to the efficiency + electrification upgrade could have instead been utilized by the power sector to generate more electricity.
 
An electricity shortfall metric was produced, defined as the difference between modeled ERCOT electricity demand and available generation, for hours in which modeled demand exceeded available generation and ERCOT requested load shed from utilities. Electricity shortfall for each building scenario is presented in Figure 10 along with ERCOT-requested load shed. Statistics on the electricity shortfall for each building scenario and ERCOT-requested load shed are presented in Table 5.

\begin{figure}[H]
\includegraphics[scale=0.45]{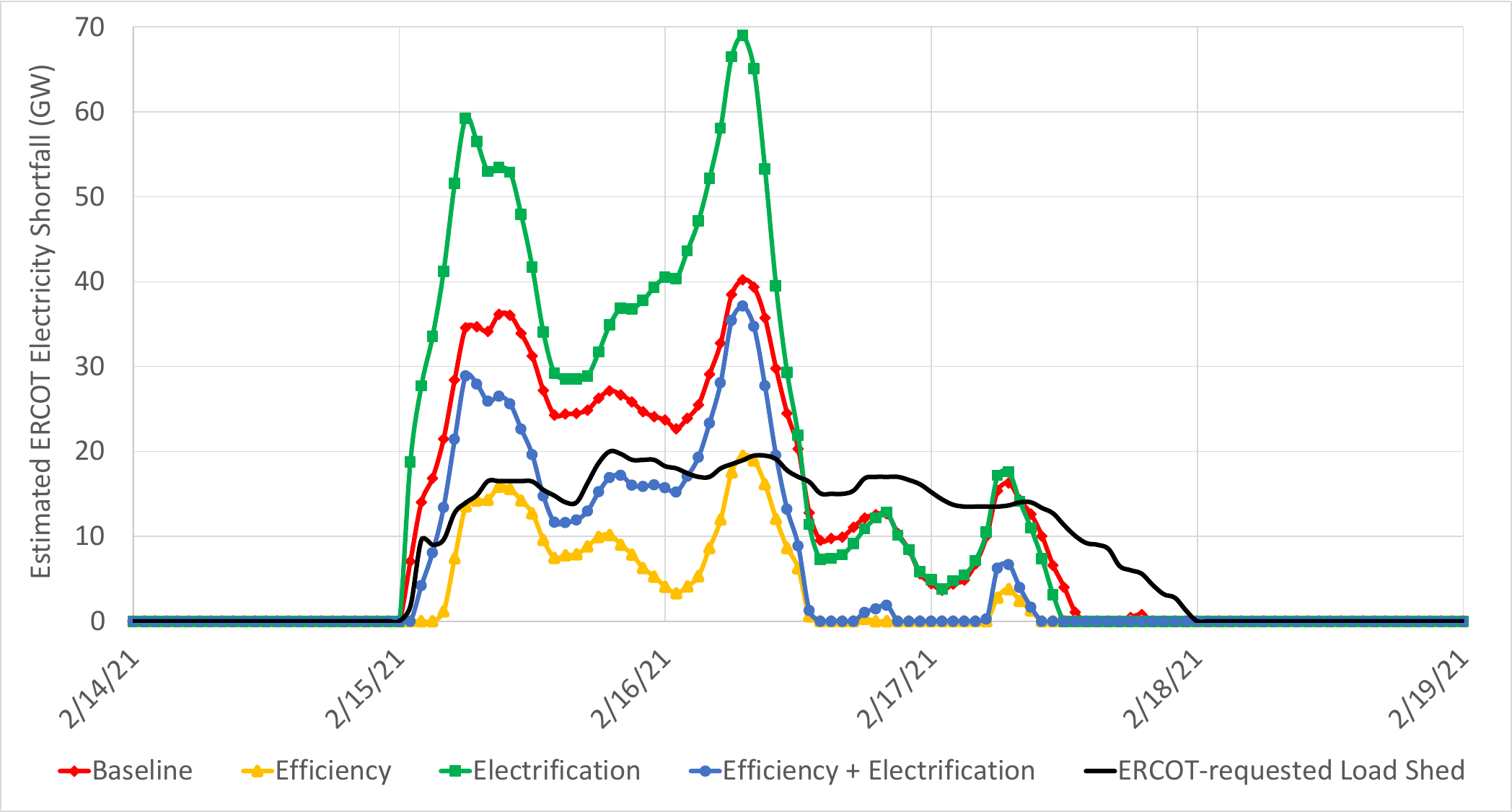}
\caption{The differences between modeled electricity demand and available generation for different upgrade scenarios indicate that the electrification upgrade package would have increased the electricity shortfall, while the efficiency and the combined efficiency + electrification upgrade packages would have reduced the electricity shortfall.}
\centering
\end{figure}

\newcolumntype{P}[1]{>{\raggedright\arraybackslash}p{#1}}
\newcolumntype{M}[1]{>{\centering\arraybackslash}m{#1}}
\newcolumntype{L}[1]{>{\raggedright\let\newline\\\arraybackslash\hspace{0pt}}m{#1}}
\begin{table}[H]
  \centering
  \begin{tabular}{|L{7.5cm}|M{1cm}|M{1cm}|M{2.5cm}|M{1cm}|} 
    \hline
    \multicolumn{1}{|>{\centering\arraybackslash}m{7.5cm}|}{Scenario}
    & \multicolumn{1}{|>{\centering\arraybackslash}m{1cm}|}{Hours}
    & \multicolumn{1}{|>{\centering\arraybackslash}m{1cm}|}{Peak (GW)}
    & \multicolumn{1}{|>{\centering\arraybackslash}m{2.5cm}|}{Peak as Percent of Load (\%)}
    & \multicolumn{1}{|>{\centering\arraybackslash}m{1cm}|}{Total (GWh)}\\ \hline\hline
    ERCOT-requested Load Shed & 71 & 20.0 & 29.4\% & 1000 \\ \hline
    Baseline - Electricity Shortfall & 63 & 40.2 & 45.4\% & 1225 \\ \hline
    Efficiency - Electricity Shortfall & 39 & 19.5 & 28.7\% & 347 \\ \hline
    Electrification - Electricity Shortfall & 59 & 69.0 & 58.7\% & 1741 \\ \hline
    Efficiency + Electrification - Electricity Shortfall & 44 & 37.1 & 43.4\% & 704 \\ \hline
  \end{tabular}
  \caption{The electrification package would have increased the magnitude of the electricity shortfall that occurred from February 15\textsuperscript{th} through February 17\textsuperscript{th}, while the efficiency and efficiency + electrification upgrade packages would have reduced the magnitude of the electricity shortfall and the number of hours where an electricity shortfall occurred.}\label{tab1}
\end{table}

\pagebreak

ERCOT-requested load shed (black line) peaked at 20.0 GW on February 15th, and totaled 1000 GWh for the entire event. The electricity shortfall for the baseline building stock had two distinct peaks at 36.1 GW and 40.2 GW on February 15\textsuperscript{th} and 16\textsuperscript{th}, respectively, and totaled 1225 GWh for the entire event. The electricity shortfall for the electrification upgrade package would have peaked at 59.2 GW and 69.0 GW on February 15\textsuperscript{th} and 16\textsuperscript{th}, respectively. Over the course of the event, the electrification upgrade would have increased the cumulative electricity shortfall by 42\% relative to the baseline scenario. The efficiency upgrade package would have reduced the number of hours where load shed occurred to 39 hours from 63 hours that occurred with the baseline scenario. The cumulative electricity shortfall for the efficiency scenario would have been 72\% lower than for the baseline scenario. With the gap between electricity supply and demand reduced this much for the efficiency scenario, it is possible that ERCOT could have maintained operating frequency by implementing less-intrusive rolling blackouts instead of extended load shed events. The electricity shortfall for the efficiency + electrification scenario would have been lower than baseline scenario electricity shortfall for every hour of the event. In total, the efficiency + electrification scenario would have reduced the total electricity shortfall by 43\% relative to the baseline scenario.

Figure 11 demonstrates how February 2021 residential and commercial electricity demand was about the same for all scenarios when population-weighted mean temperature was above 10°C but diverged at colder temperatures.

\begin{figure}[H]
\includegraphics[scale=0.85]{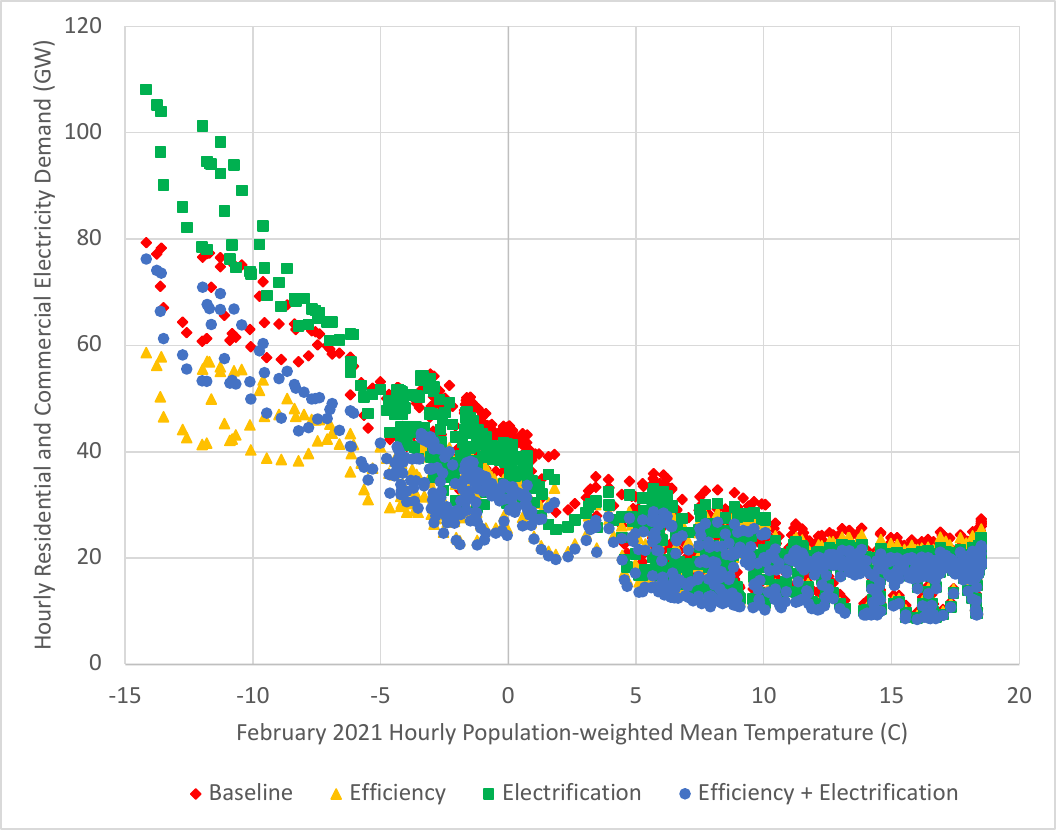}
\caption{At lower population-weighted mean temperatures, the electrification package would have increased hourly residential and commercial sector electricity demand, while the efficiency and efficiency + electrification packages would have decreased hourly residential and commercial sector electricity demand.}
\centering
\end{figure}
As hourly mean temperature decreased, residential and commercial electricity demand would have increased for all building stock scenarios. For lower heating loads (higher mean temperature), the baseline building stock scenario (red diamonds) had the highest demand. The electrification upgrade (green squares) would have slightly lowered electricity demand during high mean temperature hours due to efficiency improvements over legacy HVAC equipment. The efficiency scenario (yellow triangles) and efficiency + electrification scenario (blue circles) would have had the lowest electricity demand for high mean temperature hours due to the reduced heating load realized with improvements to building envelopes. These trends would have held until hourly mean temperature approached -5°C (23°F). At this point, diminishing ASHP efficiency and increased reliance on electric resistance heating would have caused demand for the electrification scenario to surpass baseline demand. During the coldest hours of February 2021, we estimate that peak residential and commercial electricity demand for the electrification scenario would have exceeded 100 GW. Similarly, while electricity demand for the efficiency and efficiency + electrification scenarios would have been nearly the same at moderate temperatures, as the hourly mean temperature approached -5°C (23°F), diminishing electrical heating equipment efficiency would have caused demand for the efficiency + electrification scenario to exceed demand for the efficiency scenario. Overall, both the efficiency and efficiency + electrification scenarios would have reduced residential and commercial electricity demand relative to the baseline scenario during every hour of the month, and peak residential and commercial demand for the efficiency scenario (58.6 GW) would have been about half of peak demand for the electrification scenario (108.1 GW). These trends underscore the potential for electrical heating equipment to deliver electricity savings under typical weather conditions, which could offset much of the cost of the electrification upgrade package. However, these results also highlight the potential for electrical heating to pose a power grid reliability issue if it is not deployed in conjunction with sufficient building envelope efficiency upgrades.

Figure 12 demonstrates that hourly percent savings relative to baseline demand would have been positive for all upgrade scenarios at modest temperatures.

\begin{figure}[H]
\includegraphics[scale=0.85]{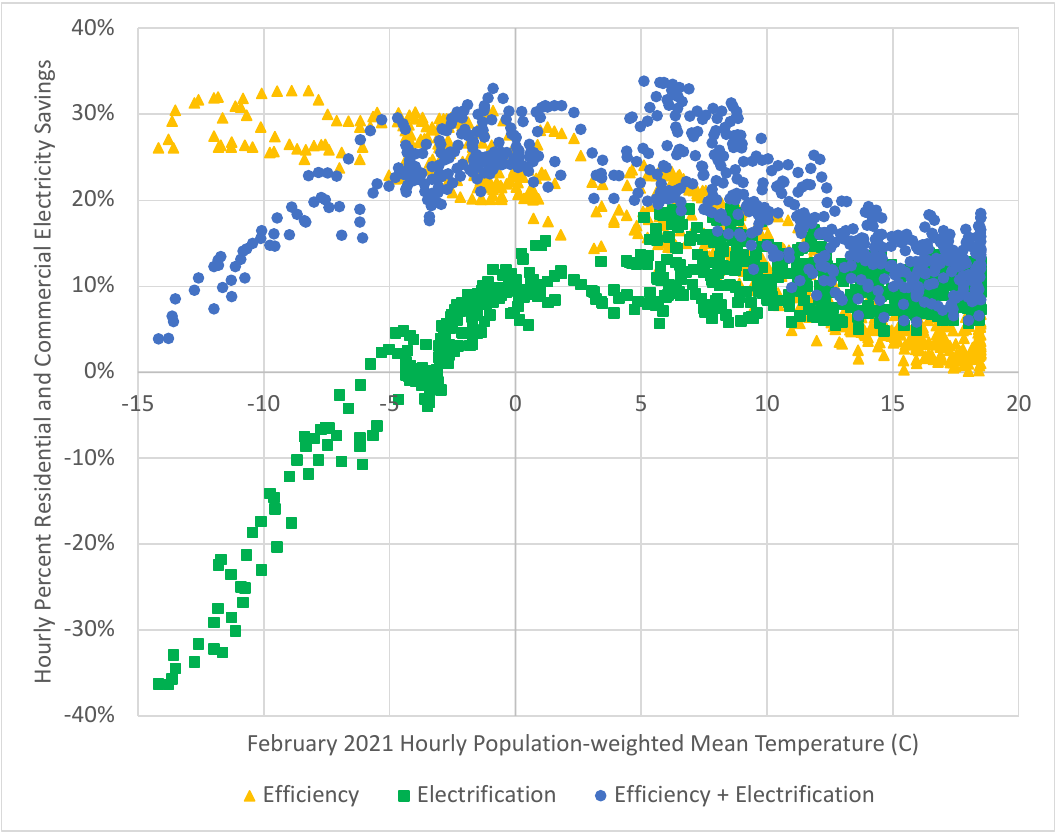}
\caption{Hourly February 2021 percent residential and commercial electricity savings for the electrification and efficiency + electrification scenarios would have been higher than for the efficiency scenario at higher population-weighted mean temperatures, but at lower mean temperatures, the savings for electrification and efficiency + electrification scenarios would have dropped below the efficiency scenario.}
\centering
\end{figure}

As hourly mean temperatures decreased towards 5°C (41°F), savings would have increased for all upgrade scenarios, with savings for the efficiency + electrification scenario (blue circles) initially exceeding savings for the efficiency scenario (yellow triangles) and the electrification scenario (green squares). However, as ambient temperatures dropped further and electrical heating efficiency diminished, savings for scenarios with ASHPs would have peaked at around 5°C (41°F) and then began to decrease. Eventually, at an hourly mean temperature of -5°C (23°F), savings for the electrification scenario would have gone negative (electrification scenario demand would have exceeded baseline scenario demand). During the coldest hours of the month, residential and commercial demand for the electrification scenario would have been 36\% greater than for the baseline scenario. Savings for the efficiency + electrification scenario would have approached zero during the coldest hours of February 2021, and would have gone negative if ambient temperatures had dropped lower. As hourly mean temperature decreased, savings for the efficiency scenario would have increased and eventually plateaued at around 30\% near an hourly mean temperature of -10°C (14°F).

While building electricity performance had a notable impact on system operations during Winter Storm Uri, the impact of building electricity performance on power grid operations the rest of the year is also important to consider.

\begin{figure}[H]
\includegraphics[scale=0.35]{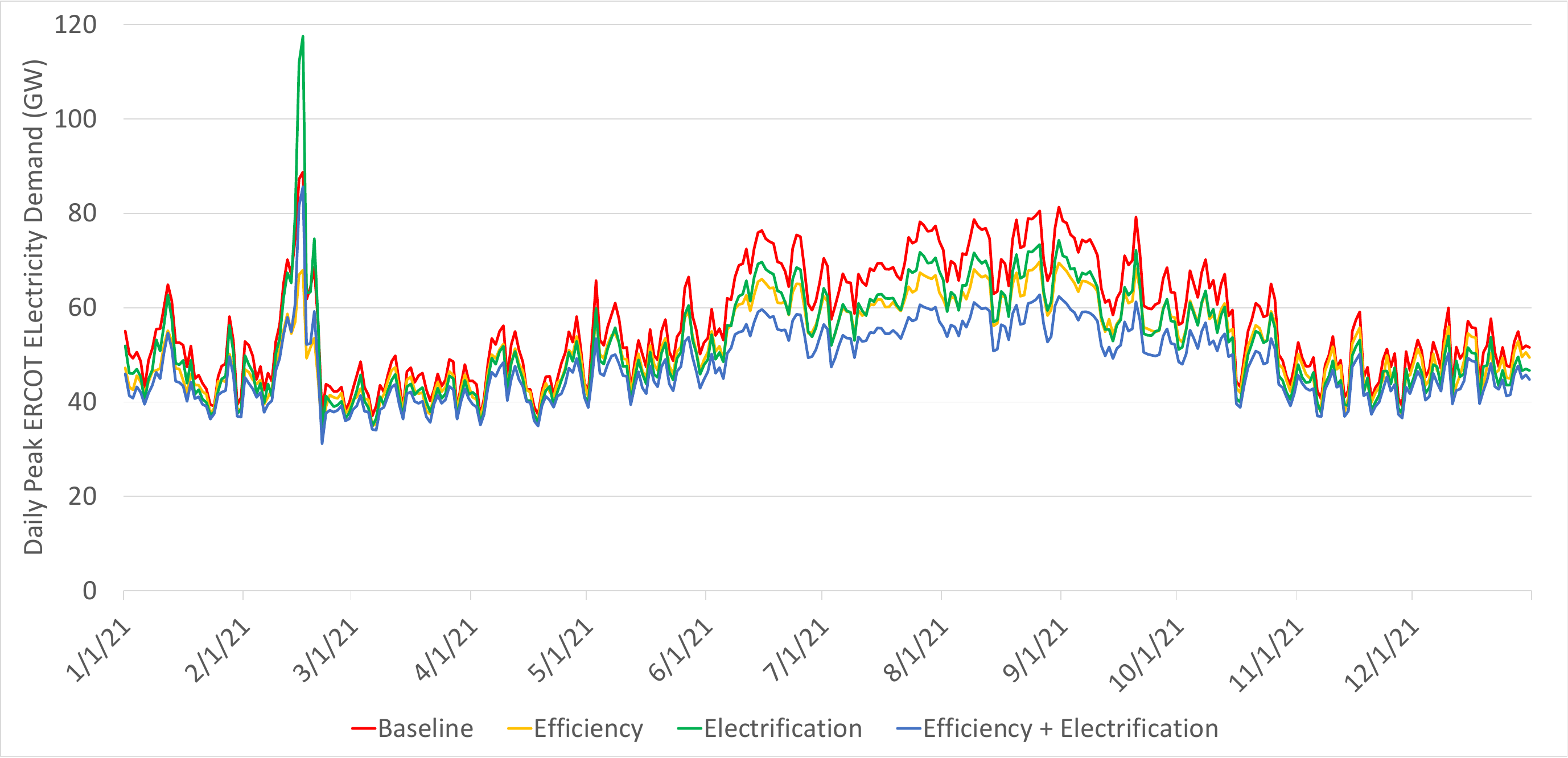}
\caption{Electrification upgrades would have reduced summer peak demand and increased winter peak demand while efficiency upgrades would have reduced both summer and winter peak demand.}
\centering
\end{figure}

As shown in Figure 13, daily peak demand for different building stock scenarios would have been similar during the moderate “shoulder months” (March – May and September – November) but would have diverged during the winter and summer. Daily peak demand for the baseline building stock ranged from 53.1 GW to 81.3 GW (difference of 28.2 GW) during the summer months (June – August) and ranged from 37.2 GW to 88.7 GW (difference of 51.5 GW) during the winter months (January, February, December). These fluctuations demonstrate the sensitivity of system peak demand to seasonal weather. If demand had been fully served during Winter Storm Uri, it would have been the first winter annual peak load in ERCOT history. This highlights a challenge associated with wintertime ERCOT operations, because while daily peak demand for 2021 summer days always fell within a 28.2 GW range, daily peak demand during the winter varied much more (51.5 GW). Furthermore, previous research has demonstrated that increasing penetration of electrical heating equipment in ERCOT is causing winter peak demand to grow more quickly and erratically year over year than summer peak demand \cite{skiles2023}.

The efficiency scenario would have reduced total 2021 ERCOT electricity consumption by 5.9\% and dampened the effect of both cold and hot temperatures on electricity demand. This is evidenced by the fact that peak demand for the efficiency scenario during Winter Storm Uri would only have been 18.4 GW higher than the average daily peak demand during the shoulder months. Conversely, baseline peak demand during Winter Storm Uri was 35.6 GW higher than the average daily peak demand during the shoulder months. The flatter, less seasonal efficiency scenario peak demand profile could potentially be served by a smaller generation fleet operating a higher percent of the time. This reduction in requisite generation capacity could lead to lower overall system costs.

The electrification upgrade package would have reduced the sensitivity of building electricity demand to hot temperatures and magnified the sensitivity of building electricity demand to very cold temperatures. Overall, the electrification package reduced total 2021 ERCOT electricity consumption by 6.8\% and lowered daily peak demand for every day of the year except during the week of Winter Storm Uri (February 14\textsuperscript{th} – 17\textsuperscript{th}, 19\textsuperscript{th} – 20\textsuperscript{th}). ASHP systems would have been particularly effective at reducing daily peak demand during summer months by delivering efficiency improvements over legacy air conditioning (space cooling) equipment. However, during Winter Storm Uri, the electrification upgrade package would have increased peak demand to 117.5 GW from 88.7 GW for the baseline scenario.  Both upgrade packages that electrified building heating would have resulted in winter demand peaks that far exceeded summer demand peaks. This suggests that ERCOT will become a regularly winter peaking grid if penetration of electric heating equipment continues to increase due to consumer preferences and government policy. The demand profile created by the electrification scenario might be particularly difficult to manage for a grid operator like ERCOT. Because while daily peak electricity demand for the electrification scenario hit 117.5 GW during Winter Storm Uri, it never exceeded 74.3 GW the rest of the year. To fully serve system demand, ERCOT would need at least 117.5 GW of available power generation connected to the grid and operational during such a winter event. However, because peak demand would never exceed 74.3 GW outside of Winter Storm Uri, at least 43.1 GW of this generation fleet would sit idle for all but a few days in 2021. Furthermore, because the temperatures seen during Winter Storm Uri rarely occur in Texas, this 43.1 GW of generation might only be dispatched for a few days in a multi-year period.

In an “energy-only” market like ERCOT, where power generators are only paid when they generate and sell power to the grid, it might be difficult for the final 43.1 GW of generation to remain financially viable. As a result, the ERCOT market might have difficulty attracting sufficient power generation capacity to serve the estimated 117.5 GW of power demanded during a Winter Storm Uri like event. An electricity market that includes capacity payments might partially alleviate this issue, however the additional capacity payments required for 117.5 GW of generation would likely increase systems costs. Coupled with the fact that the electrification scenario would reduce the amount of annually delivered electricity through which system costs are recouped, this means that the electrification scenario could result in an increase in the unit cost of electricity. This unit cost increase would cut into the cost savings that building owners might expect from a heat pump retrofit. Policy-makers should consider these market dynamics as they seek to decarbonize building heating by incentivizing heat pump adoption. These winter performance issues could be mitigated, if cold climate heat pumps are installed instead of the moderate efficiency heat pumps analyzed in this study. However, cold climate heat pumps are more expensive and might need more policy support to make them attractive to ERCOT building owners.

While the main focus of this analysis is the impact of these measures during the winter, both the efficiency and ASHP retrofits included in the efficiency + electrification upgrade package would have reduced the sensitivity of building electricity demand to hot temperatures. The ASHP retrofit would also increase the sensitivity of building electricity demand to very cold temperatures. Overall, the efficiency + electrification upgrade would have lowered total 2021 ERCOT electricity consumption by 11.9\% and reduced daily peak demand during every day of the year. This upgrade package would have been particularly effective at reducing peak summer demand, lowering peak summer demand to 62.7 GW from 81.3 GW for the baseline scenario. The efficiency + electrification upgrade would have marginally reduced peak winter demand to 85.6 GW from 88.7 GW for the baseline scenario. It is notable that heat pumps, when installed in conjunction with efficiency measures, can serve 100\% of Texas building space conditioning load and still reduce total electricity consumption and peak demand.

\section{Conclusions}
It should be noted, when resource planners consider the future of the ERCOT grid, they are really asking the question: how do we heat and cool our buildings in a reliable and cost-efficient manner? While only 32\% of 2021 ERCOT electricity consumption was dedicated to residential and commercial building space conditioning, this load increased dramatically in response to ambient temperatures to constitute most of winter and summer peak demand. Seventy-four percent of ERCOT peak electricity demand (88.7 GW) during Winter Storm Uri was comprised of residential and commercial space conditioning demand (65.9 GW). Building space conditioning could become an even larger part of winter peak demand in the future due to recent trends in electrical heating adoption and decarbonization initiatives designed to shift building heating load from natural gas infrastructure to the power grid. Similarly, building space conditioning required as much as 41.9 GW of electricity during the summer months. It is thus important to carefully examine our buildings and assess the most efficient pathways to optimize building electricity performance through a combination of electrical and mechanical upgrades. 

There is significant opportunity to improve Texas building performance as the Texas building stock has particularly inefficient envelopes and are heavily reliant on relatively inefficient electrical heating systems. An electrification scenario would have resulted in a 32.4\% increase in peak demand during Winter Storm Uri due to diminishing efficiency of electrical heating systems at cold temperatures. However, the electrification upgrade would have resulted in efficiency gains during more moderate temperatures throughout the year, leading to a 6.8\% reduction in total 2021 electricity consumption. These results demonstrate the potential for the adoption of electrical heating to save consumers energy, but also create winter reliability issues. It should be noted that the electrification scenario would have delivered this energy performance while nearly completely decarbonizing space heating in ERCOT buildings. A building efficiency scenario would have resulted in a 20.7 GW (23.4\%) decrease in peak demand during Winter Storm Uri and a 5.9\% decrease in total 2021 electricity consumption. Thus, the efficiency upgrade package would have provided the same grid service during Winter Storm Uri as 20.7 GW of additional dispatchable power generation. Peak Winter Storm Uri load shed was 40.2 GW for the baseline scenario and would have been lowered to 19.5 GW by the efficiency upgrade package. This more modest shortfall in generation could possibly have been managed with rolling blackouts instead of an extended load shed event. A combined building efficiency + electrification scenario would have resulted in a 3.1 GW (3.5\%) decrease in Winter Storm Uri peak demand and a 11.9\% decrease in total 2021 electricity consumption. All three upgrade scenarios would have lowered typical summer daily peak demand closer to typical daily peak demand during the shoulder months. This reduction in the seasonality of peak demand could reduce the amount of generation capacity required to serve system demand year-round, potentially lowering system costs. These upgrades packages delivered these benefits through static building retrofits alone. Building electricity demand response could be utilized to provide additional grid services beyond those identified by this analysis. These results demonstrate that if applied judiciously, static building sector retrofits can deliver significant savings and important reliability benefits when the ERCOT grid is most stressed.

\section{CRediT authorship contribution statement}
\textbf{Matthew J. Skiles:} Data curation; Formal Analysis; Investigation; Methodology; Visualization; Writing --- original draft, Writing --- review \& editing. \textbf{Joshua D. Rhodes:} Conceptualization; Methodology; Writing --- review \& editing. \textbf{Michael E. Webber:} Supervision; Writing --- review \& editing.

\pagebreak
\biboptions{sort&compress}
\bibliographystyle{elsarticle-num}
\bibliography{uri_electricity}

\end{document}